%% file: main.tex
\documentclass[%
 reprint,
 superscriptaddress,
 nofootinbib,
 amsmath,amssymb,
 aps,
 prd,
floatfix,
]{revtex4-2}

\usepackage{graphicx} 
\usepackage{dcolumn} 
\usepackage{bm} 
\usepackage{array} 
\usepackage{hyperref}
\usepackage[caption=false]{subfig} 
\usepackage{gensymb} 
\usepackage[mathlines]{lineno}
\usepackage{placeins}
\usepackage[
separate-uncertainty = true,
multi-part-units=single
]{siunitx}
\usepackage{xcolor}
\usepackage[version=4]{mhchem}
\usepackage[normalem]{ulem}
\usepackage{tablefootnote}
\usepackage{blindtext}
\usepackage{makecell}

\DeclareSIUnit\keVnr{keV_{nr}}
\DeclareSIUnit\keVee{keV_{ee}}

\begin{document}

\preprint{APS/123-QED}

\title{Nuclear recoil yield and fluctuation modeling in liquid xenon with DD neutron calibrations in the LUX-ZEPLIN (LZ) Experiment}
\input{authorlist}


\begin{abstract}
We report on the light and charge yields of nuclear recoil (NR) events in liquid xenon (LXe) using deuterium–deuterium (DD) neutron calibration data from the LUX-ZEPLIN (LZ) experiment.
These data informed the model used in LZ's recent search for NR events from $^8$B solar neutrinos and low-mass dark matter. 
While the mean yields inferred here are consistent with global averages, the DD data suggest that fluctuations in light and charge production are not fully described by current models of signal production in LXe. 
These fluctuations are relevant for interpreting the NR rates from $^8$B solar neutrinos and low-mass Weakly Interacting Massive Particles (WIMPs), as well as the signal shapes of higher-mass WIMPs.

\end{abstract}

\keywords{Dark Matter, Direct Detection, Xenon, Nuclear Recoil, Time Projection Chamber, Calibration}

\maketitle

\input{introduction} 
\input{DescriptionOfCalibrationSetup} 
\input{DescriptionOfCalibrationData} 
\input{SimulationAndModelingFramework}

\input{Results}

\input{Discussion}

\input{Conclusion}

\input{Acknowledgements}

\bibliographystyle{apsrev4-2}

\clearpage
\appendix
\input{Appendix}

\clearpage
\bibliography{references}

\setcounter{equation}{0}
\setcounter{figure}{0}
\setcounter{table}{0}
\makeatletter

\renewcommand{\theequation}{S\arabic{equation}}
\renewcommand{\thefigure}{S\arabic{figure}}
\renewcommand{\thetable}{S\arabic{table}}
\renewcommand{\bibnumfmt}[1]{[S#1]}
\renewcommand{\citenumfont}[1]{S#1}

\end{document}

%% file: authorlist.tex

\author{R.~Abdullah}
\affiliation{South Dakota School of Mines and Technology, Rapid City, SD 57701-3901, USA}

\author{D.S.~Akerib}
\affiliation{SLAC National Accelerator Laboratory, Menlo Park, CA 94025-7015, USA}
\affiliation{Kavli Institute for Particle Astrophysics and Cosmology, Stanford University, Stanford, CA  94305-4085 USA}

\author{A.K.~Al Musalhi}
\affiliation{University College London (UCL), Department of Physics and Astronomy, London WC1E 6BT, UK}

\author{B.J.~Almquist}
\affiliation{Brown University, Department of Physics, Providence, RI 02912-9037, USA}

\author{C.S.~Amarasinghe}
\email{chami@ucsb.edu}
\affiliation{University of California, Santa Barbara, Department of Physics, Santa Barbara, CA 93106-9530, USA}

\author{A.~Ames}
\affiliation{SLAC National Accelerator Laboratory, Menlo Park, CA 94025-7015, USA}
\affiliation{Kavli Institute for Particle Astrophysics and Cosmology, Stanford University, Stanford, CA  94305-4085 USA}

\author{T.~Anderson}
\affiliation{Lawrence Berkeley National Laboratory (LBNL), Berkeley, CA 94720-8099, USA}

\author{T.J.~Anderson}
\affiliation{SLAC National Accelerator Laboratory, Menlo Park, CA 94025-7015, USA}
\affiliation{Kavli Institute for Particle Astrophysics and Cosmology, Stanford University, Stanford, CA  94305-4085 USA}

\author{N.~Angelides}
\affiliation{University of Zurich, Department of Physics, 8057 Zurich, Switzerland}

\author{H.M.~Ara\'{u}jo}
\affiliation{Imperial College London, Physics Department, Blackett Laboratory, London SW7 2AZ, UK}
\affiliation{STFC Rutherford Appleton Laboratory (RAL), Didcot, OX11 0QX, UK}

% ...

\collaboration{The LZ Collaboration}

%% file: introduction.tex
\section{Introduction}

The nuclear recoil (NR) response in liquid xenon (LXe) defines the signal region of Weakly Interacting Massive Particles (WIMPs) in the searches carried out by LXe time projection chambers (TPCs).
NR events are also produced by the coherent elastic neutrino-nucleus scattering (CE$\nu$NS) of solar neutrinos, which now constitutes a competing signal in searches for low-mass WIMPs~\cite{pandaxcollaboration2024indicationsolar8bneutrino_PandaX4T8BfirstIndication, PhysRevLett.133.191002_XnT8BfirstIndication, akerib2025searches}.
Previous measurements of the LXe response to NR events have utilized knowledge of the scattering angle of monoenergetic neutrons to reconstruct the recoil energy~\cite{lenardo2019measurement, luxcollaboration2016lowenergy0774kevnuclear_LUXlowENRcalibration, pershing2022calibrating_LLNL_DT, lenardo2019low, plante2011new, aprile2018simultaneous_neriX, manzur2010scintillation}.
Our approach to constrain the NR yields involves comparing data to a continuous spectrum from neutron transport and detector response simulations, similar to Refs.~\cite{aalbers2025lowenergynuclearrecoilcalibration_LZYBe, xenoncollaboration2024lowenergynuclearrecoilcalibration_XnTYBe}.  

We model the LXe response to NR events in the \mbox{2--74~keV} range using the scatters of 2.45~MeV deuterium-deuterium (DD) neutrons with xenon nuclei.
Two novel mechanisms of LXe signal production are employed to describe the data. 
First, fluctuations in scintillation light production are enhanced in an energy-dependent fashion compared to prior Poisson-like expectations. 
These enhanced fluctuations can push otherwise sub-threshold events into the analysis region. 
Second, the variance in electron-ion recombination is suppressed compared to the prior binomial expectation.
Models of LXe response that lack these features are rejected by our data due to disagreements located near the 74~keV DD-endpoint. 

In this work, we characterize these mechanisms in the full 2–74 keV range and provide an NR response model used to interpret NR events in dark matter and solar neutrino searches~\cite{akerib2025searches}. The DD calibration and detector conditions are described in Sec.~\ref{sec:CalibrationApparatus}. Details on data selection, correction, and event systematics are described in Sec.~\ref{sec:calibrationDataSelection}. 
The specifications of our NR model are detailed in Sec.~\ref{sec:SimulationAndModelingFrameworkSection}, the results shown in Sec.~\ref{sec:Results}, and the implications of the new model for the detection of low-energy NR events are discussed in Sec.~\ref{sec:discussion}.

%% file: DescriptionOfCalibrationSetup.tex
\section{Calibration Apparatus}
\label{sec:CalibrationApparatus}

The LUX-ZEPLIN (LZ) dark matter experiment consists of a 7~tonne dual-phase LXe TPC and surrounding veto systems deployed underground at the Sanford Underground Research Facility~\cite{akerib2020lux}. Particle interactions in the TPC produce prompt scintillation photons (S1) and electrons that are extracted into a gas region and converted to light (S2).
Both S1 and S2 signals are detected by two arrays of photomultiplier tubes (PMTs) located at the top and bottom of the TPC. 
The electrons are drifted by a 97~V/cm electric field with gains in the S1 and S2 channels of $0.112\pm0.002$~photons detected (phd) per scintillation photon and $34.0\pm 0.9$~phd/electron, respectively.
These are the same detector conditions reported in Refs.~\cite{4dyc-z8zf_WS2024Results, akerib2025searches}.
The S1 and S2 sizes are used to reconstruct the event deposition energy, as well as discriminate between NR and electronic recoils (ER) using the S1/S2 ratio. 
The position of the interaction is reconstructed using the PMT hit map for the transverse coordinates ($x,y$)~\cite{Solovov:2011aa, Akerib_2018_LUXpositionReconstruction} and the time difference between the S1 and S2 signals for the depth coordinate ($z$).
The LZ veto systems are designed to tag multiple-scattering backgrounds and consist of an instrumented LXe skin between the TPC and the cryostat vessel and a gadolinium-loaded liquid scintillator detector submerged in water surrounding the cryostat. A detailed description of the LZ experiment can be found in Ref.~\cite{akerib2020lux}. 

LZ performs NR calibrations~\cite{Aalbers_2024_DesignImplementationPerformanceLZCalibration} using a variety of neutron sources such as AmBe, AmLi, YBe (which in Ref.~\cite{aalbers2025lowenergynuclearrecoilcalibration_LZYBe} was recently used to constrain NR yields), and an Adelphi Technologies DD109 generator~\cite{VAINIONPAA2014203_DDneutronGenerator} to measure yields and veto tagging efficiencies.
The DD generator provides unique advantages over the other sources for yield measurements, such as precise neutron generation timing, a low $\gamma$-to-neutron ratio, and multiple configurations: direct mode, hydrogen reflector mode, and deuterium reflector mode~\cite{VERBUS201768_lowEAbsoluteCalibrationNRusingDD, CaseyRhyneThesis, JeanneBangThesis}. 
This analysis uses data taken in the DD direct mode. 

\begin{figure}[h]
    \centering
    \includegraphics[width=0.95\linewidth]{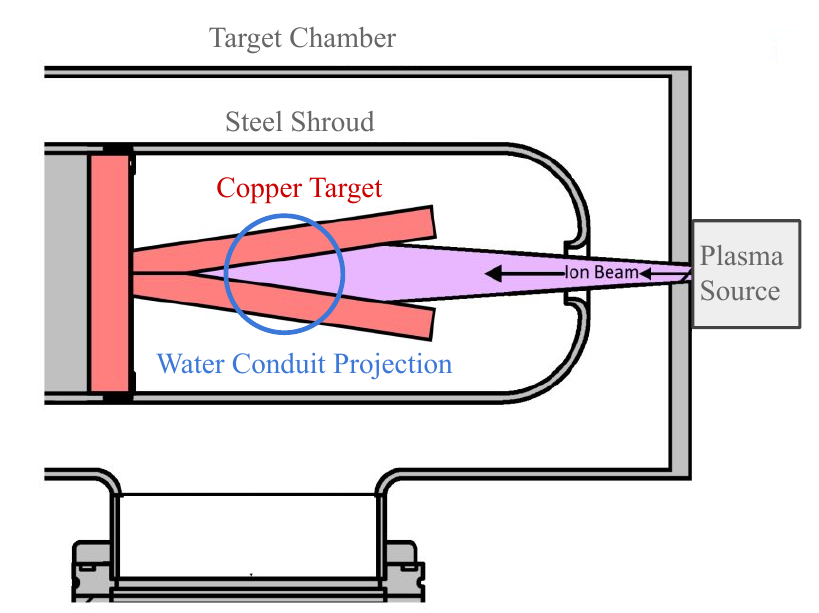} 
    \caption{Diagram of the Adelphi DD109 neutron generator modified from Ref. \cite{CaseyRhyneThesis}. The plasma source component is not to scale, while the target chamber, steel shroud, copper target, and water conduit projection are. Neutrons emitted into the page travel towards the TPC. The blue circle depicts the DD water conduit projection onto the DD head.}
    \label{fig:DDneutronGeneratorGraphic}
\end{figure}

The DD generator is a multi-stage apparatus that produces neutrons from the nuclear fusion of ionized deuterium. 
A schematic of the generator is shown in Fig.~\ref{fig:DDneutronGeneratorGraphic}. 
The process begins in the DD plasma chamber where low-pressure D${}_{2}$ gas is periodically excited into a plasma. 
The periodic ignition is achieved by using pulsed 2.469~GHz microwaves whose power and time structure are configured using feedback from a photodiode observing the plasma chamber.
Plasma generation occurs at a frequency of 150~Hz with 50~$\mu$s pulse widths to maximize neutron interactions while limiting multi-neutron pileup.
The probability of two or more neutrons reaching the TPC from the same pulse with a neutron rate of 15 n/s pulsed at 150 Hz is $\sim$~5\%. The probability of these pileup events entering our selection by having merged S1s interpreted as a single S1 by our data processing framework is less than $0.1$\%.
The ionized deuterium nuclei in each pulse are accelerated through an 81~kV potential difference to embed in a titanium-coated copper surface, where they chemically bond to form a target layer of TiD${}_{2}$.
The subsequent accelerated ions fuse with the embedded deuterium, emitting neutrons via
\begin{equation}
    {}^{2}\mathrm{H} + {}^{2}\mathrm{H} \rightarrow {}^{3}\mathrm{He} \ (0.82 \ \mathrm{MeV}) + \mathrm{n} \ (2.45 \ \mathrm{MeV}).
    \label{eqn:neutronGeneration}
\end{equation} 

Although ionized hydrogen and tritium are also produced, these charged particles are fully attenuated by the walls of the target chamber. 
Unlike $\left (\alpha, n \right )$ reactions, DD fusion is associated with low gamma emission, with a $\gamma$-to-neutron ratio of $10^{-7}$~\cite{CECIL1985411_gamma-to-neutron-ratio}.
The assembly features a 6~mm lead shield on the generator head to attenuate x-rays and $\gamma$-rays from material interactions and DD fusion~\cite{Aalbers_2024_DesignImplementationPerformanceLZCalibration}.

\begin{figure*}
    \centering
    \includegraphics[width=0.95\linewidth]{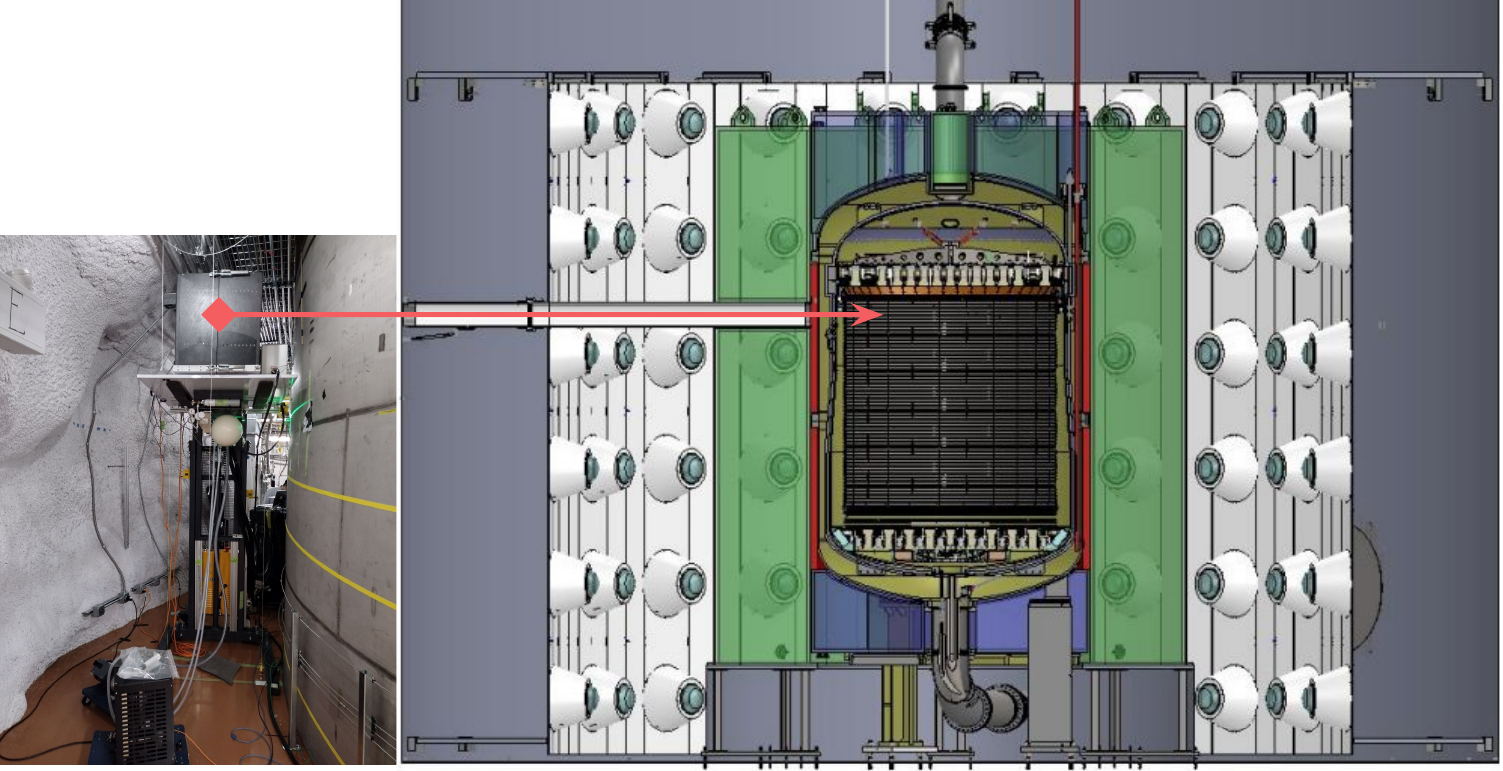}
    \caption{Location of the DD generator in the direct-mode configuration relative to the LZ detector. The photograph was taken during the calibration campaign described in this work. The black enclosure is the BPE neutron shield, within which the DD generator is housed. The white sphere is a Bonner sphere used to monitor the neutron intensity.}
    \label{fig:DD_in_LZ}
\end{figure*}

Neutrons are produced isotropically at a rate of $2\times10^{6}$~neutrons/s and monitored by a Bonner sphere positioned below the generator~\cite{MeasurementBonnerSphere}. Surrounding the generator is a borated polyethylene (BPE) neutron shield that attenuates neutrons that are off the main neutron beam being directed toward the evacuated conduit leading to the TPC.
The kinetic energy spectrum of the neutrons that exit the generator depends on the properties of the deuterium ion beam, the fusion kinematics, and scattering within the generator assembly~\cite{CaseyRhyneThesis}.
The generator is oriented to maximize the flux and purity of 2.45~MeV neutrons that enter the TPC.
The copper target is aligned with a N${}_{2}$-purged conduit that passes through the LZ water tank, as shown in Fig.~\ref{fig:DD_in_LZ}. 
The projection of the conduit onto the generator is shown in Fig.~\ref{fig:DDneutronGeneratorGraphic}.
Transport through the conduit produces a collimated beam, as neutrons outside the direct line of sight are scattered by the surrounding water. 
The resulting beam enters the detector approximately 10~cm below the LXe surface.

The photodiode monitoring the plasma provides a trigger for the LZ data acquisition system for event capture. As the neutron travel time from the DD generator and TPC is $\sim$200~ns, the photodiode response from plasma ignition provides strong coincidence between the event trigger and the corresponding neutron-induced S1 pulse. This coincidence inside the 50~$\mu$s plasma pulse allows for time-based rejection of spurious backgrounds, which is capitalized on in this work.

%% file: DescriptionOfCalibrationData.tex
\section{Data Corrections and Selection}
\label{sec:calibrationDataSelection}

Eighty-eight live-hours of DD data were acquired for this analysis and corrected for systematic variations in the LZ TPC. PMT gains were calibrated every twelve hours using the detector response to single electrons extracted from the LXe~\cite{NEVES201013_CoimbraVUVmethod}.
The single photoelectron (SPE) efficiency and the probability of afterpulsing were measured for each PMT using LED calibrations~\cite{AALBERS2024169712_LZDAQFADR}. Incident vacuum ultraviolet (VUV) photons can cause double photoelectron emission (DPE)~\cite{Faham_2015_wavelengthDependentDPE, ResponseOfPhotomultiplierTubesToXenonScintillationLight}, where the DPE probability for each of the PMTs was measured using ${}^{83\mathrm{m}}$Kr data and incorporated into the PMT gain measurements. Spatial corrections for light collection efficiency (LCE) and electron extraction efficiency were derived from ${}^{222}$Rn alpha decays and applied to the S1 and S2 signals. 
The S2 signal was further corrected for the finite electron lifetime ($\sim15$~ms in the DD run) measured using ${}^{222}$Rn and ${}^{218}$Po alpha decays. 
The fully corrected areas of the S1 and S2 signals are denoted by S1$c$ and S2$c$. 

\begin{figure}
    \centering
    \includegraphics[width=0.95\linewidth]{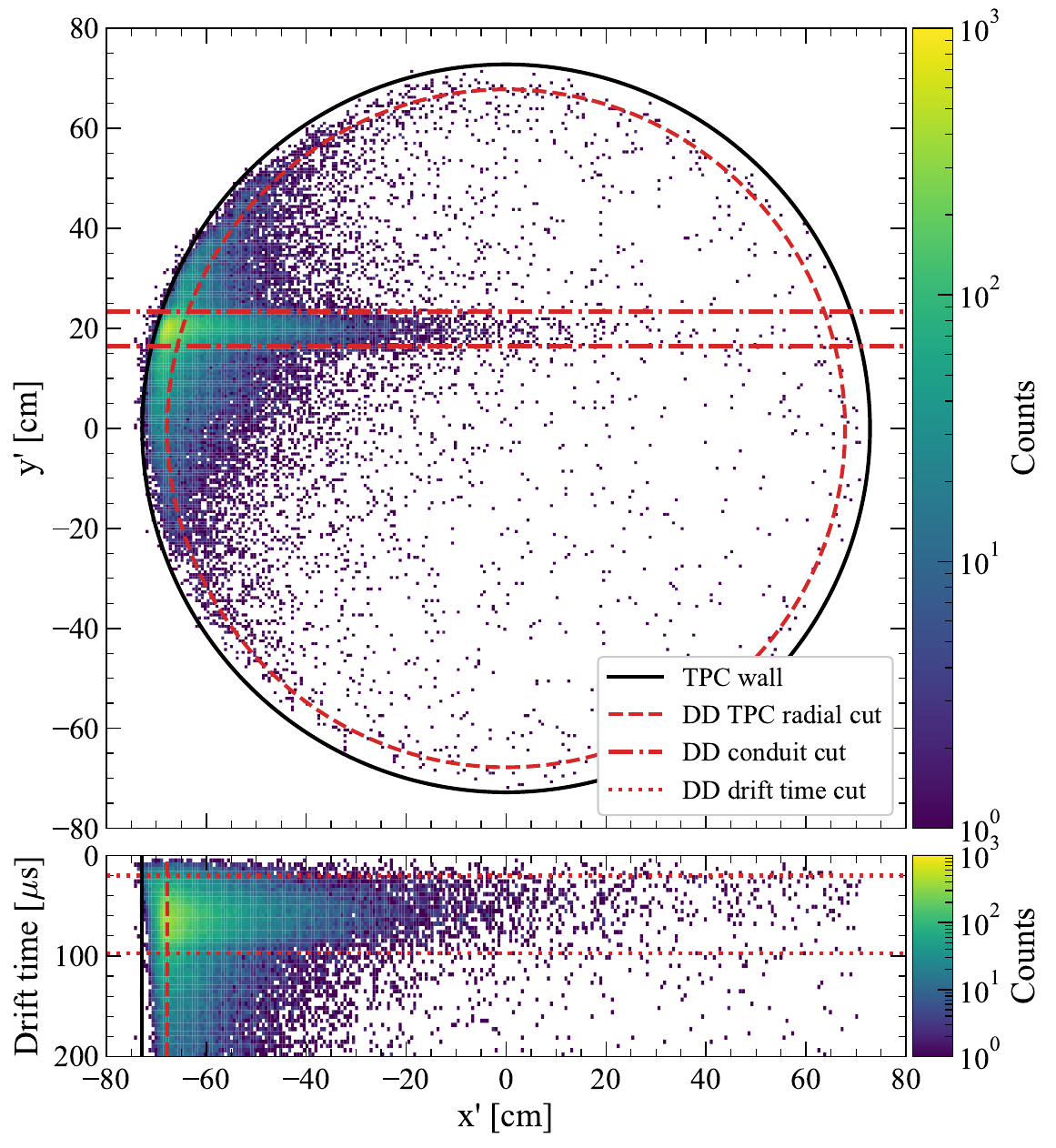}
    \caption{Position of events that pass all analysis and data quality cuts without spatial cuts applied. The dashed red and solid black circles depict the TPC radial cut and physical TPC wall, respectively. Events that exist outside of the physical TPC wall are from poor position resolution. The red dotted and dot-dashed lines depict the drift time and conduit cuts, respectively.
The $(x', y')$ coordinates reflect a counterclockwise rotation of ${\theta}_{\mathrm{rot}} = 8.0\degree$ of the standard TPC $(x, y)$ coordinates to make the position selection parallel to the horizontal axis.}
    \label{fig:DD_pos_dist}
\end{figure}

Single-scatter (SS) events with S1 pulses formed by at least a 3-fold PMT coincidence are selected for this analysis.
The region of interest is further defined as 2~$<$~S1$c$~$<$~162~phd and 220~$<$~S2$c$~$<$~14390~phd. 
Events with S1 pulses outside a 50~$\mu$s window around the plasma trigger are excluded. 
Scatters that arise from collimated neutrons are selected with position cuts that restrict events to be within the beam axis. 
A 5~cm cut from the PTFE wall is applied to reduce the contributions of neutrons that scatter in the PTFE and LXe skin to the final data.
This standoff distance is comparable to the average 5.4~cm used in Ref.~\cite{akerib2025searches}.
Events are accepted regardless of whether they were tagged by the veto system. 
Further data quality cuts are applied based on the S2 width to drift time relationship, the quality of the position reconstruction, and the likelihood of neutrons interacting with gaseous xenon (GXe). 
Figure~\ref{fig:DD_pos_dist} shows the position of neutron events after all cuts except for position selections.
Tritiated methane (CH${}_{3}$T) and AmBe calibration data were used to characterize the S1 and S2 cut efficiencies, respectively, on the DD data and are implemented for the modeling framework. 

A true multi-scatter event with vertices at similar drift times may have merged S2 pulses and be mistakenly identified as SS. 
If not accounted for, these unresolved multi-scatters (UMS) bias the yield result since they on average produce more charge than true SS at the same reconstructed energy. 
Thus, UMSs are targeted with cuts based on the width of the S2 pulse and anisotropies in the PMT hit pattern. 
UMSs that still remain after the application of these cuts are modeled in simulations, and their rate is estimated with a data-driven analysis.
Measurements of vertical separation between the S2 pulses of resolved multiple scatters are extrapolated to give an UMS rate estimate of around~5\% in the full dataset after all cuts.
The UMS fraction is ultimately treated as a nuisance parameter in the fit.

\begin{figure}
    \centering
    \includegraphics[width=0.95\linewidth]{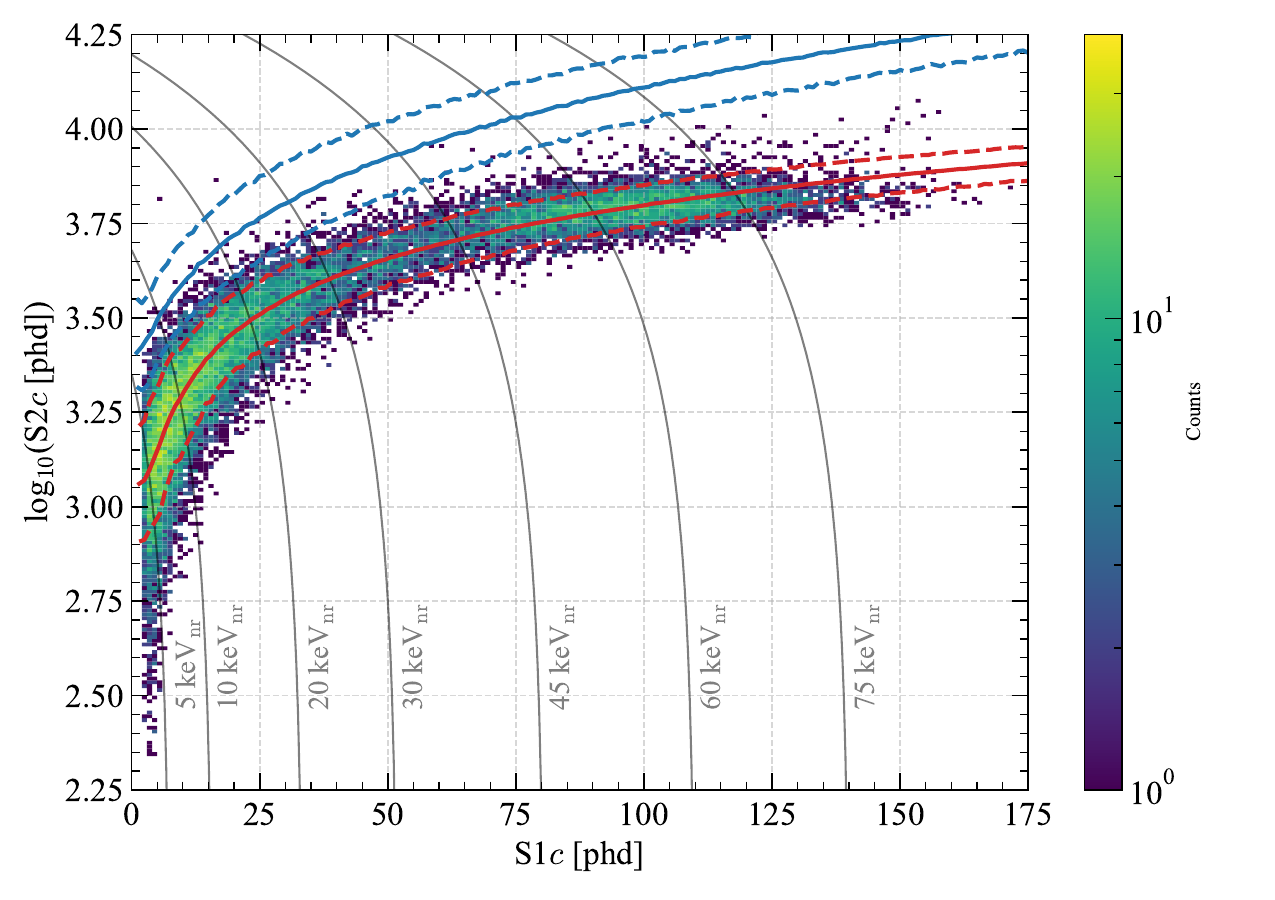}
    \caption{Distribution of the DD data after all selections have been applied in log${}_{10}$(S2$c$) vs S1$c$ space. 
    The red (blue) solid lines represent the medians of NR (ER) events simulated using flat energy spectra, and fit with skew-gaussians in 1~S1$c$~phd bins. 
    The red (blue) dashed lines are the NR (ER) 10\% and 90\% quantiles. The simulated NR events use the model from this work, while the simulated ER events use the same model as Ref.~\cite{4dyc-z8zf_WS2024Results}. Energy lines depict the nuclear recoil energy (denoted as keV ${}_{\mathrm{nr}}$) using the NR model from this work.}
    \label{fig:DD_S1_S2}
\end{figure}

After all selections, 20371 events remain and are shown in log${}_{10}$(S2$c$) vs S1$c$ space in Fig.~\ref{fig:DD_S1_S2}. The ER band uses the same parameterization as the ER model in Ref.~\cite{4dyc-z8zf_WS2024Results}, while the NR band uses the updated NR model described in this work. Both bands were simulated using a flat energy spectrum with the Noble Element Simulation Technique (\textsc{nest})~\cite{M_Szydagis_2011_NESTcomprehensivemodel, 10.3389-fdest.2024.1480975_NESTmodelsExhaustiveComparison} in the LZ detector. 

%% file: SimulationAndModelingFramework.tex
\section{Simulation and Modeling} 
\label{sec:SimulationAndModelingFrameworkSection}

This analysis extracts the mean yields and fluctuations of NR events using DD neutron scattering data to constrain a LXe response model.
The first step of the analysis is the simulation of neutron transport through the conduit and detector (Sec.~\ref{subsec:neutron_simulation}).
This stage depends only on the neutron kinematics, cross sections, and geometry, and gives a set of NR energy deposits with their positions. 
Second, these deposits are converted into S1$c$ and S2$c$ using a
\textsc{nest}-based parameterization of the LXe response
(Sec.~\ref{subsec:lxe_micro}), followed by the LZ detector response and
selection efficiencies (Sec.~\ref{subsec:observable}).
The model parameters are inferred with a Markov Chain Monte Carlo (MCMC; Sec.~\ref{sec:Results}), which proposes sets of model parameters, simulates (S1$c$, S2$c$) templates for each, and compares them with the data to sample the posterior distribution. Some default parameter values from \textsc{nest~v2.4.0}~\cite{szydagis_2023_8215927_NEST_2-4-0} were used as priors for the MCMC (described further in Sec.~\ref{sec:Results}), while  \textsc{nest~v2.4.5~beta}~\cite{szydagis_2025_17851406_NEST_2-4-5-beta} features changes in the parameterization as a result from this work. 
Unless specified otherwise, this work refers to \textsc{nest~v2.4.0}.

\subsection{Neutron simulation}
\label{subsec:neutron_simulation}

The simulation framework generates NR events produced by elastically scattering neutrons on a LXe target.
The angle of elastic scattering in the center-of-mass frame $\theta_\mathrm{CM}$ determines the NR energy as 
\begin{equation}\label{eq:elastic}
    E_\mathrm{NR} = \frac{4m_\mathrm{n}m_\mathrm{Xe}}{(m_\mathrm{n} +m_\mathrm{Xe})^2} \frac{(1-\cos \theta_\mathrm{CM})}{2} E_\mathrm{n},
\end{equation}
where $m_\mathrm{n}$ is the mass of a neutron, $m_\mathrm{Xe}$ is the mass of the xenon nucleus, and $E_\mathrm{n}$ is the neutron kinetic energy.  
The SS recoil energy spectrum depends on two components highlighted in Eq.~\ref{eq:elastic}. The first is the distribution of $\theta_\mathrm{CM}$, governed by the n--Xe elastic angular differential cross-sections, and the second is the kinetic energy $E_\mathrm{n}$ of the neutrons entering the TPC. The maximum possible NR energy in xenon, corresponding to a scattering angle defined by $\cos\theta_\mathrm{CM}=-1$,~is~$\sim$3\% of $E_\mathrm{n}$. 

\begin{figure}
    \centering
    \includegraphics[width=0.95\linewidth]{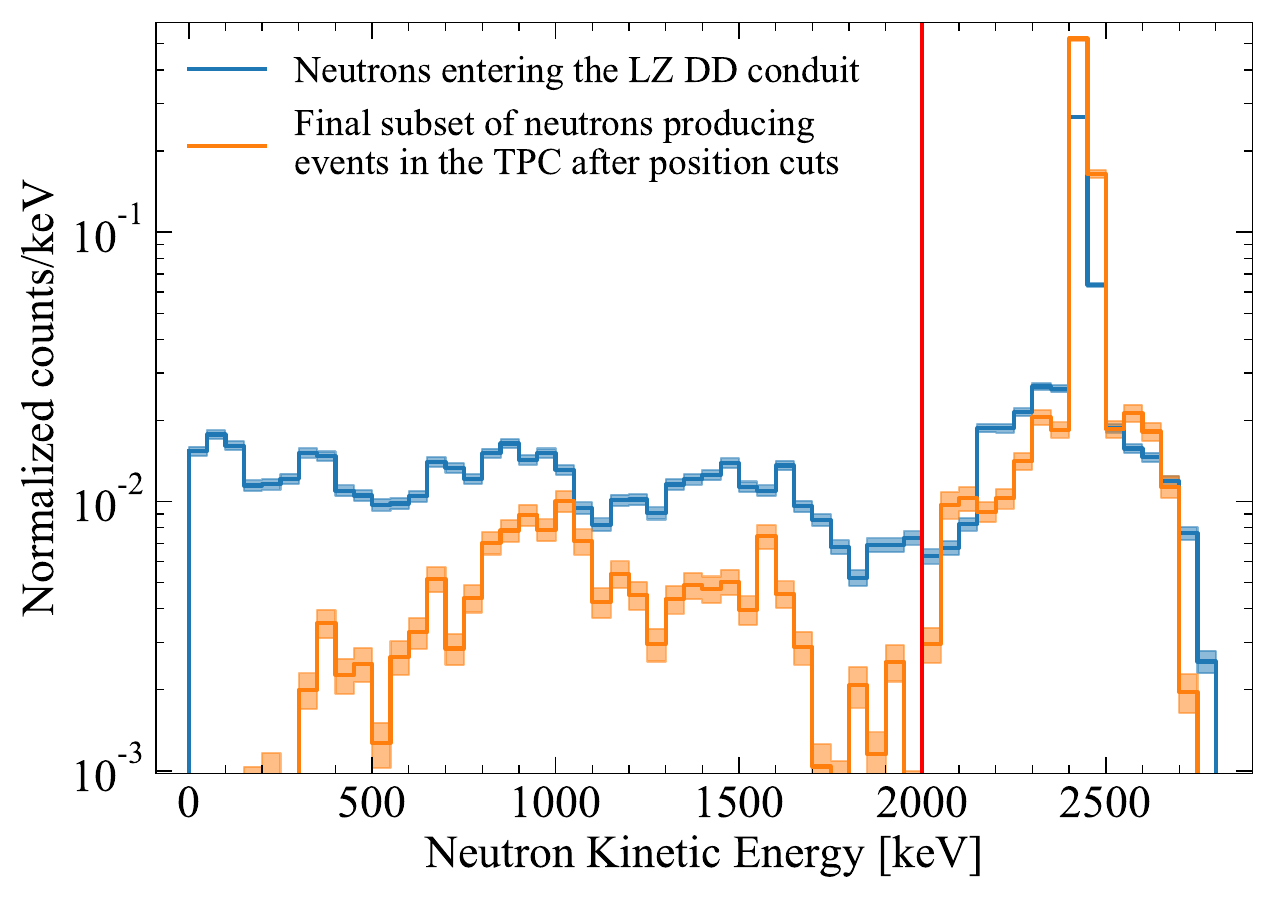}
    \caption{The simulated kinetic energy spectrum of neutrons entering the LZ conduit is shown in blue. 
    This includes neutrons taking a direct path to the TPC via the conduit, and those which are not aligned with the conduit.
    The kinetic energy spectrum of neutrons that survive transport through the conduit to make detectable NR events in the TPC (and further pass the position selection outlined in Sec.~\ref{sec:calibrationDataSelection}) is shown in orange. Both distributions are individually normalized to unity. The shaded bands are the statistical errors for each bin. The red line in the figure indicates a neutron kinetic energy of 2 MeV.
}
    \label{fig:neutrin_KE_spectrum}
\end{figure}

Neutron transport is simulated using the \textsc{QGSP\_BIC\_HP} physics list of the GEANT4-based framework~\cite{AGOSTINELLI2003250_Geant4SimulationToolkit} within the BACCARAT~\cite{AKERIB2021102480_SimulationOfEventsLZ} simulation toolkit designed by LZ\footnote{The as-built LZ geometry was implemented in simulation, with the exception of neoprene on either side of the DD conduit. With the analysis selections, particularly the position selection, the effects of neoprene in the simulation are subdominant to existing systematics on the shape of the recoil energy spectrum.}.
Figure~\ref{fig:neutrin_KE_spectrum} shows the simulated neutron kinetic-energy spectrum at the LZ neutron-conduit entrance, together with the subset that enters the TPC and produces detectable NR events after position selections. The sub-2~MeV population is produced mainly by inelastic scatters in the DD generator materials, dominated by the copper target~\cite{CaseyRhyneThesis}.
The main 2.45~MeV peak has an intrinsic width of $\sim$1\%, driven mainly by the angular spread in the deuterium ion beam, while D--D angular kinematics and elastic scatters populate the broader 2.1--2.8~MeV off-peak region~\cite{CaseyRhyneThesis}.

NR energy deposits from each interaction point within the TPC are used within the \textsc{nest}-based framework to compute observable S1 and S2 signals, accounting for light collection efficiencies, reconstruction efficiencies, and detector resolution.
We replicate the data selection by choosing simulated events that (i) produce only one S1 and one S2 signal (i.e. a SS event) and (ii) satisfy the position selection in the TPC.
These selections also allow UMS events to enter the dataset.
We isolate the UMS population in simulation by identifying interaction sites that would result in a single merged S2 according to the radial and vertical resolution of the detector. While the vertical resolution is constant ($\sim$1~mm) throughout the detector, the radial simulation depends on the S2 size, ranging from $\sim$2~cm for S2~$\leq$~200~phd and $\sim$0.9~cm for S2~$\geq$~1400~phd. The position of UMS events is calculated as a S2-area-weighted average of the individual positions of the interaction vertices. We then apply the position selection to obtain the final set of events selected in the simulation. 
This yields an expected UMS fraction of around 5\%, which is consistent with the data-driven estimate.

\begin{figure}
    \centering
    \includegraphics[width=1.0\linewidth]{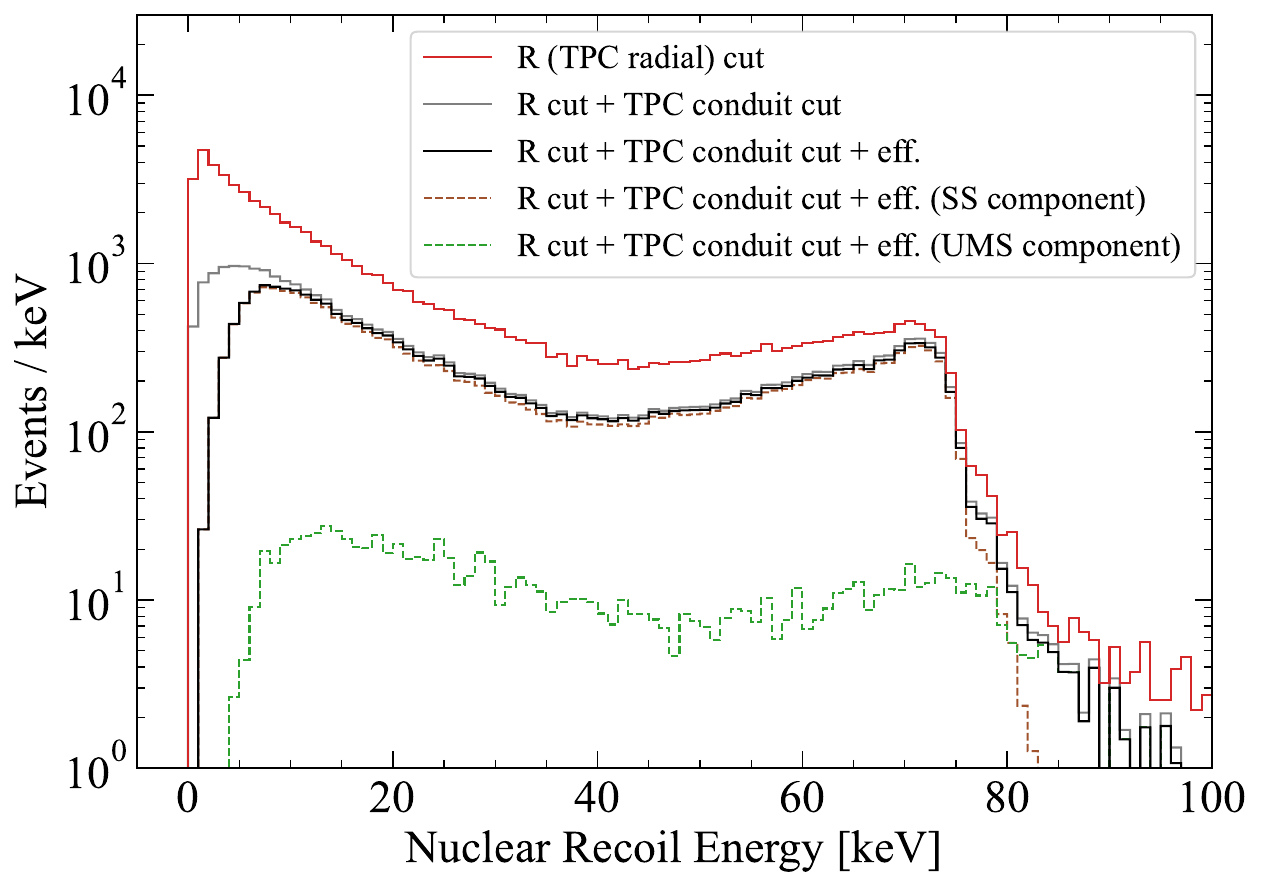}
    \caption{
    Simulated nuclear recoil energy spectra after applying sequential analysis criteria. The initial spectrum after the radial cut is shown in red, with the addition of the conduit position selection shown in light gray. The final spectrum (solid black) incorporates all data quality acceptances, reconstruction efficiencies, and analysis cuts outlined in Sec.~\ref{sec:calibrationDataSelection}. 
    This final result is decomposed into its SS (dashed brown) and UMS (dashed green) components. All curves are normalized such that the total integral of the black curve matches the observed event count in data. This spectrum has been obtained using the ENDF/B-VII.1~\cite{CHADWICK20112887_ENDF7-1} neutron data library, the default library for GEANT4.
    }
\label{fig:recoil_energy_spectrum}
\end{figure}

The recoil energy spectrum of all simulated events after sequentially applying cuts and efficiencies is shown in Fig.~\ref{fig:recoil_energy_spectrum}; this spectrum is used to compute the S1$c$ and S2$c$ for each set of \textsc{nest} parameters sampled in the fit. 
At $E_\mathrm{n}=2.45$~MeV the reduced neutron wavelength ($\approx3$~fm) is comparable to the xenon nuclear radius ($\approx6$~fm), so elastic scattering is mainly diffractive~\cite{Feshbach:1954zz, Koning:2003zz}, producing the general shape of the recoil energy distribution with a forward-peaked structure and a diffraction minimum at around 40~keV. 
Nearly isotropic compound-elastic scattering also contributes, mainly at higher recoil energies where the diffractive cross section is small, and competes with the inelastic channels available at this $E_\mathrm{n}$~\cite{Hauser:1952zz, Peters:2017xtp}. 
The position selection removes low-energy events that are caused by neutrons attenuated during conduit transport, revealing an underlying roll-off below 10~keV.
This roll-off is due to the low efficiency for the SS criterion to accept forward scattering neutrons in the beam axis, which are likely to scatter again in the LXe and be classified as multiple scatter or reconstructed as higher-energy UMS. 
The UMS events contribute significantly to the recoil spectrum above 80~keV, where the high-energy elevated S2 tail is used to constrain the UMS fraction.

\begin{figure}
    \centering
    \includegraphics[width=0.98\linewidth]{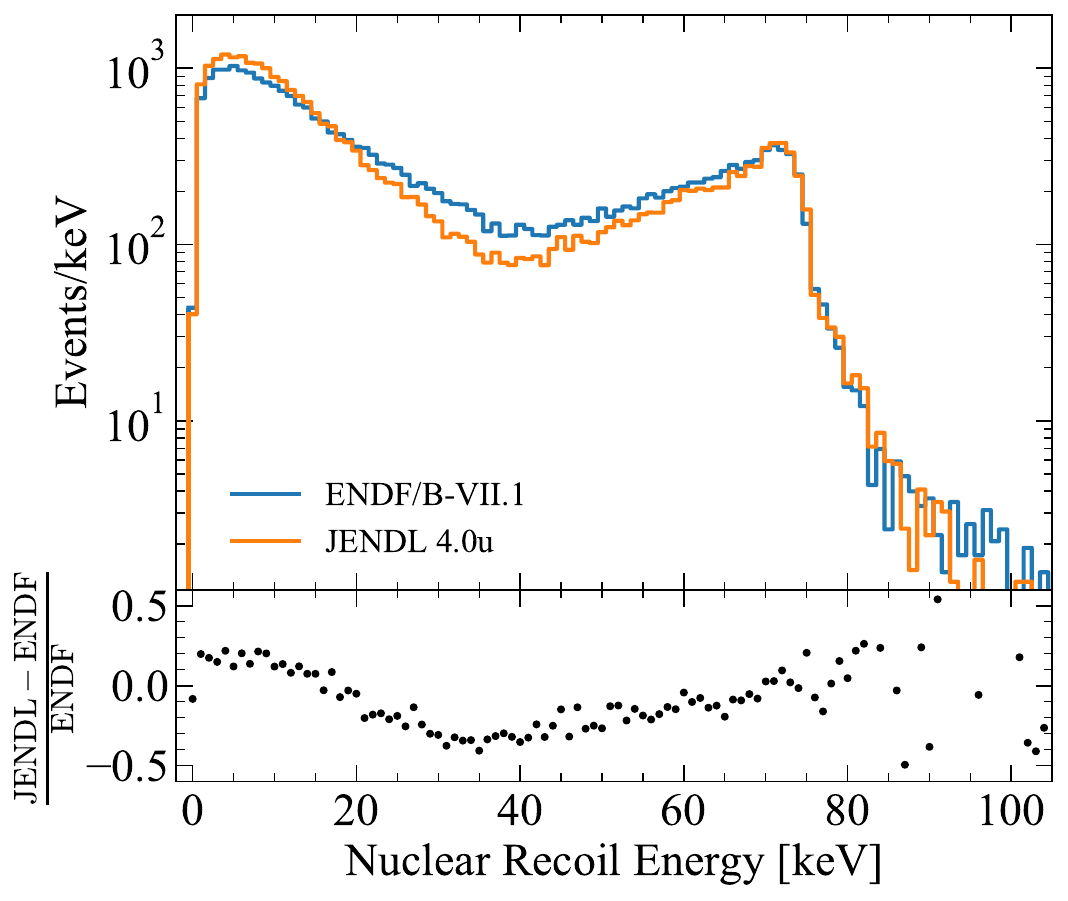}
    \caption{The top figure shows the nuclear recoil energy spectrum produced using either the ENDF/B-VII.1 library or the JENDL4.0u library for the neutron elastic cross-sections as the blue and orange lines, respectively. The bottom panel represents the fractional difference between the recoil spectra obtained using these nuclear libraries. Note that these spectra do correspond to the light gray distribution in Fig.~\ref{fig:recoil_energy_spectrum}.
    }
    \label{fig:recoil_spectrum_g4ndl_jendl}
\end{figure}

The calibration relies on the shape of the simulated recoil energy spectrum, so we consider the variation among commonly used evaluated cross section libraries~\cite{CHADWICK20112887_ENDF7-1,JENDL_4-0u}.
The libraries rely on nuclear calculations that are constrained by existing measurements; the data and inputs to these calculations are described in Ref.~\cite{Capote:2009qqf}.
A linear combination of the spectra generated by the JENDL4.0u~\cite{JENDL_4-0u} and ENDF/B-VII.1~\cite{CHADWICK20112887_ENDF7-1} libraries is fed into the LXe detector response model, with the relative fraction left as a free parameter in the fit.
These spectra are shown in Fig.~\ref{fig:recoil_spectrum_g4ndl_jendl}.

\subsection{Liquid xenon response}
\label{subsec:lxe_micro}

We have so far described \textsc{nest} as a means of converting energy deposits into observables. 
Here we describe the parameterization of the LXe response, i.e. the quantities that the fit constrains. 
Energy deposits in liquid xenon are partitioned into scintillation light, ionization charge, and heat. 
The heat channel is not detectable in LZ and is not used further in this analysis. 
The detectable light and charge signals are modeled using the treatment in \textsc{nest}~\cite{szydagis_2023_8215927_NEST_2-4-0}. 
This section also describes the modifications made to the fluctuation model of the \textsc{nest} NR signal framework to better describe the ($\mathrm{S1}c$, $\mathrm{S2}c$) distributions observed in the DD data.

The mean total number of signal quanta generated from energy deposits in NR interactions follows a power law dependence
\begin{equation}
  \langle N_{\mathrm{q}}\rangle = \alpha\,\Bigl(\tfrac{E_{\mathrm{NR}}}{1~\mathrm{keV}}\Bigr)^{\beta},
  \label{eq:Nq_powerlaw}
\end{equation}
with $\alpha$ and $\beta$ being determined from high energy NR calibrations. This is the sum of the total number of excitons ($N_{\rm ex}$) and electron-ion pairs ($N_{\rm i}$) generated from the interaction at the pre-recombination stage. The excitons along with the recombined electron-ion pairs determine the observable S1 size, while the remaining un-recombined electrons generate the S2 signal. This recombination depends on both the drift field and the deposited energy. The dependence of the charge yield on the drift field is modeled as a power law in \textsc{nest}. The DD data was collected at a constant drift field of 97 V/cm. Under the condition of having a uniform and constant drift field, the power law exponent ($\delta$) and coefficient ($\gamma$) become fully degenerate and the variation of only one is required. Therefore, we fixed $\delta$ to a constant value (the same as the default mean of \textsc{nest}), while $\gamma$ was obtained from the fitting procedure. The best fit value and the prior of $\gamma$, along with other \textsc{nest} parameters discussed herein, are shown in Table~\ref{tab:nest_params}. Although the above configuration necessitates a direct anti-correlation between the light and charge yields in the detector, Eq. \ref{eq:Nq_powerlaw} remains valid in an energy range of $E_{\mathrm{NR}} > 10$~keV. The anti-correlation is broken at low energies, resulting in the simultaneous roll-off of the light and charge yields. The parameters $\zeta$ and $\eta$ govern the low energy shape of the charge yield, while $\theta$ and $\iota$ control the same for the light yield~\cite{szydagis_2023_8215927_NEST_2-4-0}.

The fluctuations in the observed S1 and S2 arise from (i) primary-quanta statistics, (ii) recombination, and (iii) detector effects (light collection efficiency, electron extraction efficiency, electron lifetime, and electroluminescence gain).
Fluctuations in pre-recombination primary quanta are expressed via Fano-like factors for ions \(F_{\mathrm{i}}\) and excitons \(F_{\mathrm{ex}}\):
\begin{align}
  \mathrm{Var}(N_{\mathrm{i}}) &= F_{\mathrm{i}}\,\langle N_{\mathrm{i}}\rangle, &
  \mathrm{Var}(N_{\mathrm{ex}}) &= F_{\mathrm{ex}}\,\langle N_{\mathrm{ex}}\rangle .
\end{align}
The Fano-like factor for ions, $F_{\mathrm{i}}$, affects the width of the S1 (through recombination) and S2 distributions, while $F_{\mathrm{ex}}$ only controls the width of the S1 distribution. We emphasize that $N_\mathrm{ex}$ and $N_\mathrm{i}$ are not directly measurable quantities, but are modified by recombination to produce the number of potentially detectable photons and electrons to be measured as S1 and S2 signals, respectively.
We observe enhanced S1 and suppressed S2 fluctuations in the data compared to the \textsc{nest~v2.4.0} model\footnote{The DD-based tuning used for the LZ WS2022 result reduced \(F_{\mathrm{ex}}\) from 1 to 0.4~\cite{PhysRevLett.131.041002_WS2022Results}, as reflected in the LZ WS2022 detector header file, although no energy dependence was implemented at that time. The value \(F_{\mathrm{ex}}=0.4\) was later made standard in the \textsc{nest} NR model in \textsc{nest~v2.3.11}~\cite{szydagis_2022_7061832_NEST_2-3-11}.}.
Described below are the changes to the \textsc{nest} framework required to reproduce DD data.

In \textsc{nest~v2.4.0}, the recombination variance for both NRs and ERs is modeled as the sum of a binomial term and a non-binomial contribution whose magnitude is controlled by the prefactor \(A\), as given in Eq.~(8) of Ref.~\cite{10.3389-fdest.2024.1480975_NESTmodelsExhaustiveComparison}. For NRs, the default value of this prefactor is \(A=0.04\). To allow for sub-binomial behavior favored by data, we set the non-binomial prefactor \(A=0\) and introduce a dimensionless parameter \(\lambda\) that scales the binomial recombination variance.
Conditioned on a given \(N_{\mathrm{i}}\), the electron variance after recombination is
\begin{equation}
  \mathrm{Var}(N_{\mathrm{e}} \mid N_{\mathrm{i}}) \;=\; \lambda\,N_{\mathrm{i}}\,\hat r\,(1-\hat r)\,\,,
  \label{eq:lambda_conditional}
\end{equation}
where $\hat{r}$ is the recombination probability. The choice \(\lambda=1\) recovers the binomial limit, whereas \(\lambda>1\) and \(\lambda<1\) correspond to supra-binomial and sub-binomial recombination variance, respectively. Because \(N_{\mathrm{i}}\) fluctuates, the marginal variance follows from the law of total variance:
\begin{equation}
  \mathrm{Var}(N_{\mathrm{e}})
  =
  \underbrace{\lambda\,\hat r(1-\hat r)\,
  \langle N_{\mathrm{i}}\rangle}_{
  \mathbb{E}[\mathrm{Var}(N_{\mathrm{e}}\mid N_{\mathrm{i}})]}
  +
  \underbrace{(1-\hat r)^2\,
  \mathrm{Var}(N_{\mathrm{i}})}_{
  \mathrm{Var}(\mathbb{E}[N_{\mathrm{e}}\mid N_{\mathrm{i}}])}\,,
  \label{eq:lambda_marginal}
\end{equation}
The first term represents the recombination variance, whereas the second represents the ionization variance. Suppressing the ionization variance alone is insufficient: even in the limiting case \(F_{\mathrm{i}}=0\), the binomial recombination variance produces an S2 distribution broader than that observed in the data. Therefore, the recombination variance was allowed to be fit sub-binomial as \(\lambda<1\).

We capture the observed broadening of the S1 distribution by allowing supra-Poisson exciton statistics, such that
\begin{equation}
  F_{\mathrm{ex}} \;\equiv\; \frac{\mathrm{Var}(N_{\mathrm{ex}})}{\langle N_{\mathrm{ex}}\rangle} \;>\; 1 .
\end{equation}
The NR S1 distribution width empirically requires \(F_{\mathrm{ex}}\) to increase with the expected exciton count, so an $F_{\mathrm{ex}}$ dependence on $N_{\mathrm{ex}}$ is adopted as
\begin{equation}
  F_{\mathrm{ex}}
  \;=\;
  f_{\mathrm{ex}}^{(0)}
  \;+\;
  f_{\mathrm{ex}}^{(1)}\,\langle N_{\mathrm{ex}}\rangle ,
  \label{eq:Fex_linear}
\end{equation}
with \(f_{\mathrm{ex}}^{(0)}\) and \(f_{\mathrm{ex}}^{(1)}\) floated in the fit (implicitly energy dependent through \(\langle N_{\mathrm{ex}}\rangle\)). The data do not require a second-order term in \(F_{\mathrm{ex}}\).
Together with the sub-binomial recombination fluctuation, this parameterization reproduces the observed S1 and S2 variances across the full DD energy range.

\subsection{Observable signal modeling}
\label{subsec:observable}

After simulating energy deposits, generating quanta, and applying a position-dependent LCE, we keep events that satisfy a 3-fold PMT coincidence. 
Before fitting simulation to data, we adjust the simulation to account for inefficiencies that impact the S1 pulse and single-scatter reconstruction. 
First, ambient SPEs from elevated fluorescence due to high event activity may pile-up with neutron-induced two-fold S1 pulses and be promoted into the selection. 
A merging probability of 5\% was measured based on the ambient SPE rate prior to the neutron S1 pulses, where the pulse area and coincidence contribution was implemented in the simulation. 
Additionally, the pulse finding and classification software occasionally fragments real S1 pulses during reconstruction, leading to area and coincidence loss. 
This effect was measured separately using CH${}_{3}$T calibration data and included in the DD simulations. 
These efficiencies are applied event by event in the simulation, at the level of detected photons and reconstructed pulses, and depend on event position through the light collection efficiency. 
They are therefore specific to the spatial distribution of the DD events.
Pulse-based cut acceptances of the DD data are also applied to the simulation as shown in Fig.~\ref{fig:recoil_energy_spectrum}.
These cut acceptances were calculated using dedicated CH${}_{3}$T and AmBe calibration datasets. 

\subsection{Data-driven background modeling}
\label{subsec:backgroundmodel}

Unlike the UMS and sub-2~MeV neutron-population models, the background of accidental coincidences between isolated S1 and S2 pulses is modeled entirely from data. Accidental events are characterized using two sidebands of the DD neutron data: unphysical drift time (UDT) events with a reconstructed drift time \(>1100~\mu\mathrm{s}\) (beyond the maximum TPC drift time), and events outside the prompt plasma-to-S1 timing window. Both sidebands are studied in the low-energy region \(2<\mathrm{S1}c<10~\mathrm{phd}\) and \(220<\mathrm{S2}c<1000~\mathrm{phd}\), as events outside this range have negligible accidental contributions. The UDT sample is used to construct the accidental spatial model and gives a conservative upper limit of 3.7\% of events in the defined low-energy region. The 2.45 MeV neutron travel time of $\sim$200~ns allowed DD plasma ignition to effectively coincide with the neutron-induced S1 pulse. Since event capture was triggered on the plasma response, this means that the event trigger coincides with the S1 pulse within a pulse window of 50~$\mu$s. Using the off-peak sidebands of the S1-to-trigger timing ($\mathrm{S1}_{\mathrm{time}}$), the accidental population is measured to be less than 1.1\% of the main peak between $-5 < \mathrm{S1}_{\mathrm{time}} < 45~\mu\mathrm{s}$. 

In the model used in the LZ $^8$B observation~\cite{akerib2025searches}, the number of accidental events was fixed at the data-driven upper-limit, corresponding to a 3.7\% fraction in a low-energy region defined by $2<\mathrm{S1}c<10$~phd and $220<\mathrm{S2}c<1000$~phd. In the present work, the accidental rate is treated as a fitted parameter with an upper limit of 3.7\%.
Except for this change and a minor correction to the position distribution of DD events, the setup of the models is identical.
The impact on the $^8$B result was cross-checked and determined to be negligible. This cross-check is discussed in Appendix~\ref{sec:accidentals}.

%% file: Results.tex
\section{Results}
\label{sec:Results}

The light- and charge-yield models are constrained by sampling a binned likelihood with a Markov Chain Monte Carlo (MCMC). Similar approaches were performed in Refs.~\cite{7317823_globalAnalysisLightChargeYields, xenoncollaboration2024lowenergynuclearrecoilcalibration_XnTYBe}. The MCMC framework is provided by the python package emcee~\cite{Foreman-Mackey_2013_emceeHammer} using a Gaussian differential evolution proposal~\cite{Nelson_2014}. 
The likelihood evaluation is performed as follows: For a given set of parameters proposed by the MCMC algorithm, a 2D template in S1$c$--S2$c$ is constructed using the simulation procedure described in the previous section. 
A Dirichlet multinomial likelihood~\cite{MacKay2003-information_theory} is computed for the binned dataset, using the simulated template as the truth value.
We focus the constraint on the relative distribution of events rather than the absolute rate, and match the total number of simulated events in each template to the size of the observed dataset~\cite{MacKay2003-information_theory}. 
Because this normalization is primarily anchored by the high-statistics region near the endpoint and the main NR band, the neutron interaction cross sections extrapolate the counts into lower energy regimes where it would otherwise be difficult to constrain the rate. 

We implement flat priors for most of the parameters, and use Gaussian priors for the $\theta$, $\iota$, $\zeta$ and $\eta$ \textsc{nest} parameters that control the mean yield response below around 5~keV. A cross-check confirmed that gaussian priors on all \textsc{nest} parameters did not significantly impact the posteriors with the exception of the aforementioned four low-energy parameters.
The Gaussian priors originate from \textsc{nest} fits to global data~\cite{szydagis_2023_8215927_NEST_2-4-0, 10.3389-fdest.2024.1480975_NESTmodelsExhaustiveComparison}, and are used due to low statistics near the analysis threshold.
All prior and posterior ranges are shown in Table~\ref{tab:nest_params}.

\begin{table*}
    \centering
    \caption{Default and best-fit model parameters for the \textsc{nest} NR mean yield, fluctuation, response model, and nuisance parameters as defined in the text. In the column labeled priors, the values in the square brackets define upper and lower boundaries of a flat prior for the parameter. The Gaussian priors for the low energy roll-off parameters ($\zeta$, $\eta$, $\theta$ and $\iota$) are represented by their respective mean and standard deviation. \label{tab:nest_params}}
    
    \begin{ruledtabular}
    \begin{tabular}{lccc}
    NR Yield Parameter & \textsc{nest~v2.4.0} Default & LZ Prior & LZ Posterior \\
    \hline
    $\alpha$ (power law base term for number of quanta) & $11.0^{+2.0}_{-0.5}$ & [8.0, 17.0] & $11.3\pm1.1$\\
    $\beta$ (power law exponential term for number of quanta) & 1.10 $\pm$ 0.05 & [0.80, 1.25] & $1.091^{+0.024}_{-0.022}$ \\
    $\gamma$ (power law base for electric field dependence) & 0.0480 $\pm$ 0.0021 & [0.0150, 0.0650] & $0.0527^{+0.0005}_{-0.0006}$ \\
    $\delta$ (power law exponent for electric field dependence) & $-0.0533 \pm 0.0068$ & -0.0533 (fixed) & -0.0533 (fixed) \\
    $\epsilon$ [keV] (reshaping parameter for charge yield) & $12.6^{+3.4}_{-2.9}$ & [2.0, 30.0] & $8.3^{+1.8}_{-1.6}$ \\
    $\zeta$ [keV] (charge yield roll-off energy) & 0.3 $\pm$ 0.1 & $0.3 \pm 0.1$ & $0.46 \pm 0.15$ \\
    $\eta$ (charge yield roll-off shape) & 2 $\pm$ 1&  $2.0 \pm 1.0$ & $1.1^{+0.5}_{-0.2}$ \\
    $\theta$ [keV] (light yield roll-off energy) & 0.30 $\pm$ 0.05 & $0.3 \pm 0.05$ & $0.30 \pm 0.09$ \\
    $\iota$ (light yield roll-off shape) & 2.0 $\pm$ 0.5 & $2.0 \pm 0.5$ & $2.3^{+0.8}_{-0.7}$ \\
    \hline
    \hline
    \\
    \hline
    \hline
    NR Fluctuations Parameter & \textsc{nest~v2.4.0} Default & LZ Prior & LZ Posterior \\
    \hline
    $F_{\mathrm{i}}$ (constant) & 0.4 & [0.001, 2.000] & $1.1 \pm 0.4$ \\
    $f_{\mathrm{ex}}^{(0)}$ ($F_{\mathrm{ex}}$ constant term) & 0.4 & [0.001, 8.000] & $1.1^{+1.4}_{-0.8}$ \\
    $f_{\mathrm{ex}}^{(1)}$ ($F_{\mathrm{ex}}$ linear coefficient) & - & [0.0, 0.1] & $0.022 \pm 0.006$\\
    $A$ (non-binomial recombination variance prefactor) & 0.04 & 0 (fixed) & 0 (fixed) \\
    $\lambda$ (recombination variance)  & 1 & [0.01, 1.20] & $0.31^{+0.22}_{-0.19}$ \\    
    \hline
    \hline
    \\
    \hline
    \hline
    \multicolumn{2}{l}{Nuisance Parameter} & LZ Prior & LZ Posterior \\
    \hline
    \multicolumn{2}{l}{JENDL/(ENDF+JENDL) fraction} & [0.0, 1.0] & $0.63^{+0.11}_{-0.16}$ \\
    \multicolumn{2}{l}{KE $<2$~MeV neutron fraction} & [0.0, 0.3] & $0.11 \pm 0.03$ \\
    \multicolumn{2}{l}{Unresolved multi-scatter background fraction} & [0.0, 0.2] & $0.050^{+0.012}_{-0.011}$ \\
    \multicolumn{2}{l}{
      \makecell[l]{
        Accidental coincidence background fraction\\
        $2<\mathrm{S1}c<10$ phd, $220<\mathrm{S2}c<1000$ phd
      }
    } & [0, 0.037] & {$0.019^{+0.012}_{-0.013}$} \\
    \end{tabular}
    \end{ruledtabular}
\end{table*}

\begin{figure*}
    \centering
    \includegraphics[width=0.95\textwidth]{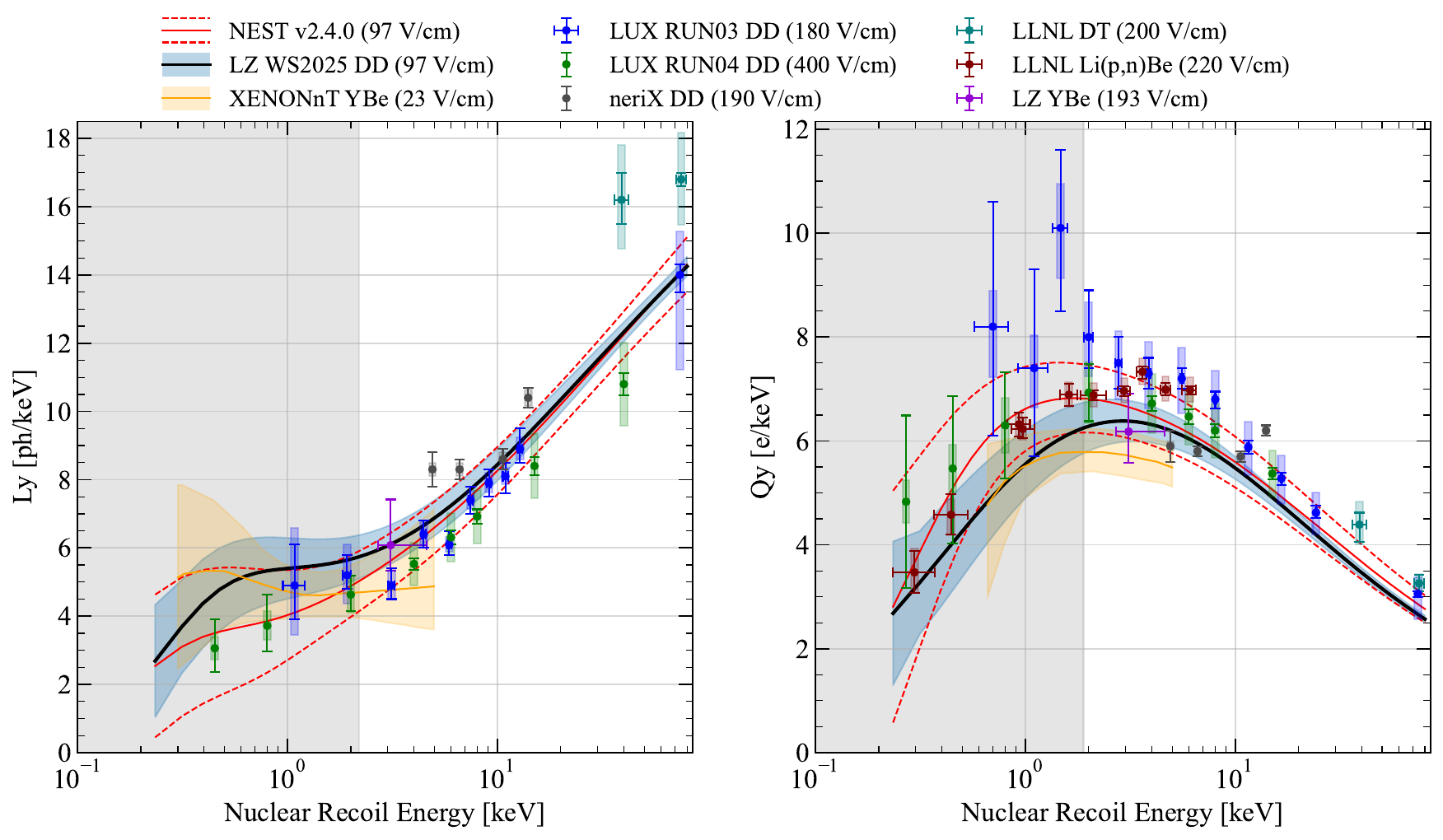}
    \caption{Light (left) and charge (right) yields corresponding to the best fit \textsc{nest} parameters as the solid black lines with $1\sigma$ uncertainty in shaded blue.  The world-fit to data from the \textsc{nest~v2.4.0} default model are shown in solid (dashed) red for the mean ($1\sigma$) light and charge yields. Other data points are direct measurements from other experiments \cite{luxcollaboration2016lowenergy0774kevnuclear_LUXlowENRcalibration, aalbers2025lowenergynuclearrecoilcalibration_LZYBe, aprile2018simultaneous_neriX, pershing2022calibrating_LLNL_DT, xenoncollaboration2024lowenergynuclearrecoilcalibration_XnTYBe,  PhysRevLett.134.061002_NRcalibrationSubKevEnergies, lenardo2019low, lenardo2019measurement}. Vertical error bars show the uncertainties each publication gives with its measurements; shaded boxes show the quadrature sum of the systematic uncertainties they quote separately. These measurements are shown at their own drift fields (legend). 
    The yields inferred in this analysis with the modification in the fluctuations modeling are consistent with \textsc{nest~v2.4.0} at all recoil energies. Model uncertainty is derived from the uncertainty on the \textsc{nest} parameters using the MCMC posterior and dominates at low energy. At higher energies, the $\sim$1.7\% and $\sim$2.6\% uncertainty of $g_1$ and $g_2$, respectively, begin to dominate.
    The shaded gray region marks the energy range in which this calibration is insensitive. 
    }
    \label{fig:best-fit-yield}
\end{figure*}

The fitted light and charge yields are consistent with \textsc{nest~v2.4.0} within \(1\sigma\) across the full recoil-energy range, as shown in Fig.~\ref{fig:best-fit-yield}. Below 5~keV nuclear recoil energy, the posterior median light yield is higher and the charge yield is lower than in \textsc{nest}'s global fit, but both remain statistically consistent with the nominal model. 
The external measurements in Fig.~\ref{fig:best-fit-yield} were taken at drift fields from 23 to 400~V/cm. A higher field suppresses recombination, which raises the charge yield and lowers the light yield. The inferred external measurements' charge yields have improved agreement when scaling to match 97~V/cm in this work using \textsc{nest~v2.4.0}. 
In the case of XENONnT YBe (23~V/cm)~\cite{xenoncollaboration2024lowenergynuclearrecoilcalibration_XnTYBe}, the inferred charge yield would increase by 8\% while the LUX Run04 measurement (400~V/cm)~\cite{PhysRevLett.134.061002_NRcalibrationSubKevEnergies} would decrease by about 7\%. The inferred scaled light yields would similarly vary anticorrelated to the charge yield due to this field dependence.

The output of the MCMC procedure is a set of samples that have converged to the underlying distribution of the posterior \cite{MacKay2003-information_theory}. 
Summary statistics such as the median and percentiles of the posterior distribution are used as estimators of the values of the parameters. These quantities are reported in Table~\ref{tab:nest_params}, and their correlations in Appendix~\ref{app:corr}.
Average light and charge yields with their corresponding uncertainties computed from the ensemble of samples in the MCMC output are shown in Fig.~\ref{fig:best-fit-yield}. 
The quoted yield uncertainties include the posterior spread of the \textsc{nest} parameters as well as the independent uncertainties on $g_1$ and $g_2$.
The yield parameterization scheme used in the LZ $^8$B result~\cite{akerib2025searches} is presented in Appendix~\ref{app:eigen}.

\begin{figure}
    \centering
    \includegraphics[width=0.95\linewidth]{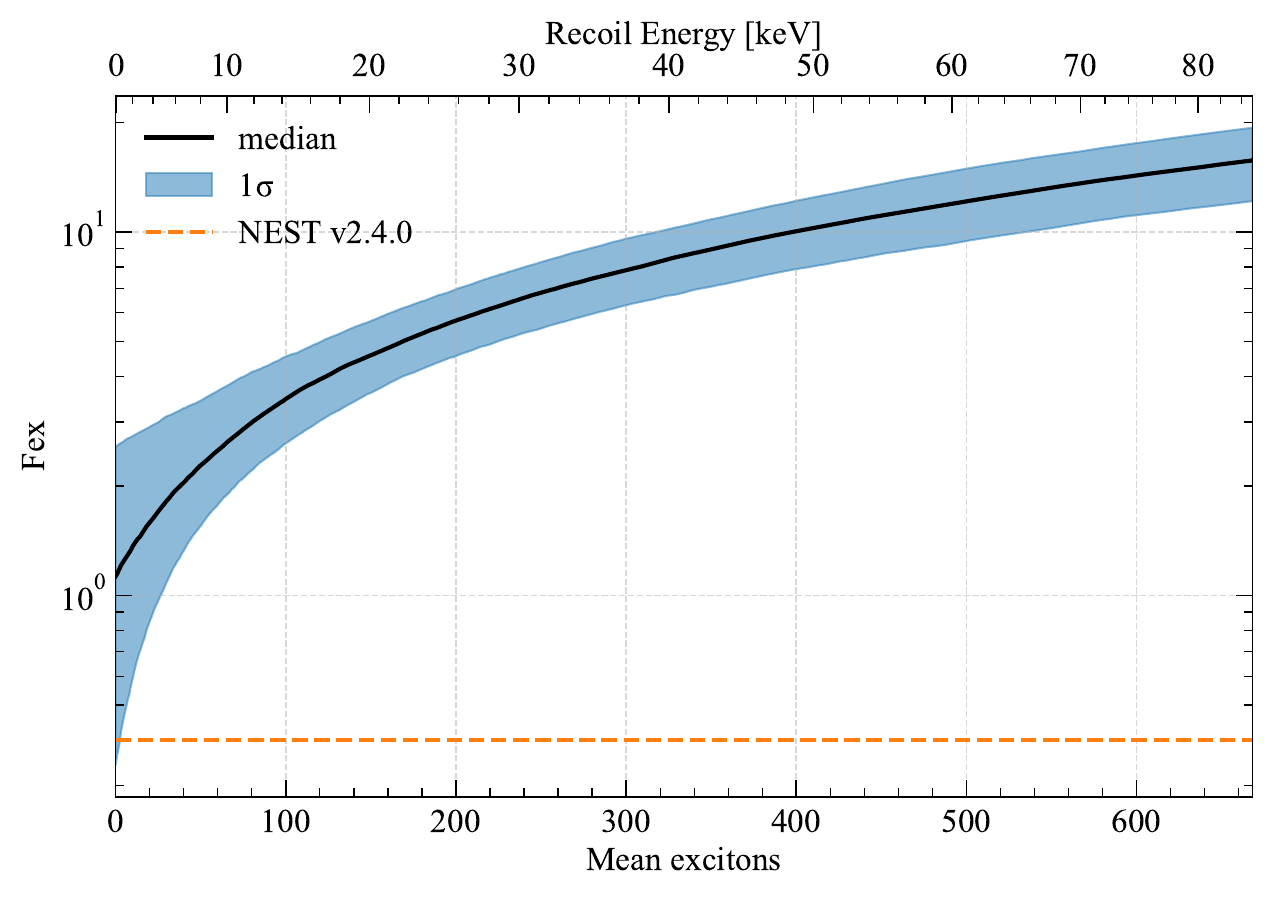}
    \caption{Best-fit $F_{\mathrm{ex}}$ as a function of both number of excitons and the corresponding nuclear recoil energy. The blue band indicates the model uncertainty obtained from the MCMC chains. The dashed orange line indicates the default \textsc{nest~v2.4.0} $F_{\mathrm{ex}}$ of 0.4 that is constant with energy. Note that $F_{\mathrm{ex}}$ dependence is based on the number of pre-recombination excitons.}
    \label{fig:fex_vs_nex}
\end{figure}

As discussed in Sec.~\ref{subsec:lxe_micro}, fitting the DD data requires modifications to the modeling of fluctuations in the LXe response. The fit requires an $F_{\mathrm{ex}}$ that increases with energy. The linear parametrization of the number of pre-recombination excitons described in Eq.~\ref{eq:Fex_linear} yields best-fit values of \(f_{\mathrm{ex}}^{(0)} = 1.1^{+1.4}_{-0.8}\) and \(f_{\mathrm{ex}}^{(1)} = 0.022 \pm 0.006\). Figure~\ref{fig:fex_vs_nex} illustrates the conversion of these values into an $F_{\mathrm{ex}}$ for different numbers of excitons and the corresponding mean energy. 
The change in recombination fluctuation discussed in Eq.~\ref{eq:lambda_conditional} produces a best-fit \(\lambda = 0.31^{+0.22}_{-0.19}\).

\begin{figure*}[t]
  \centering
    \includegraphics[width=0.49\textwidth]{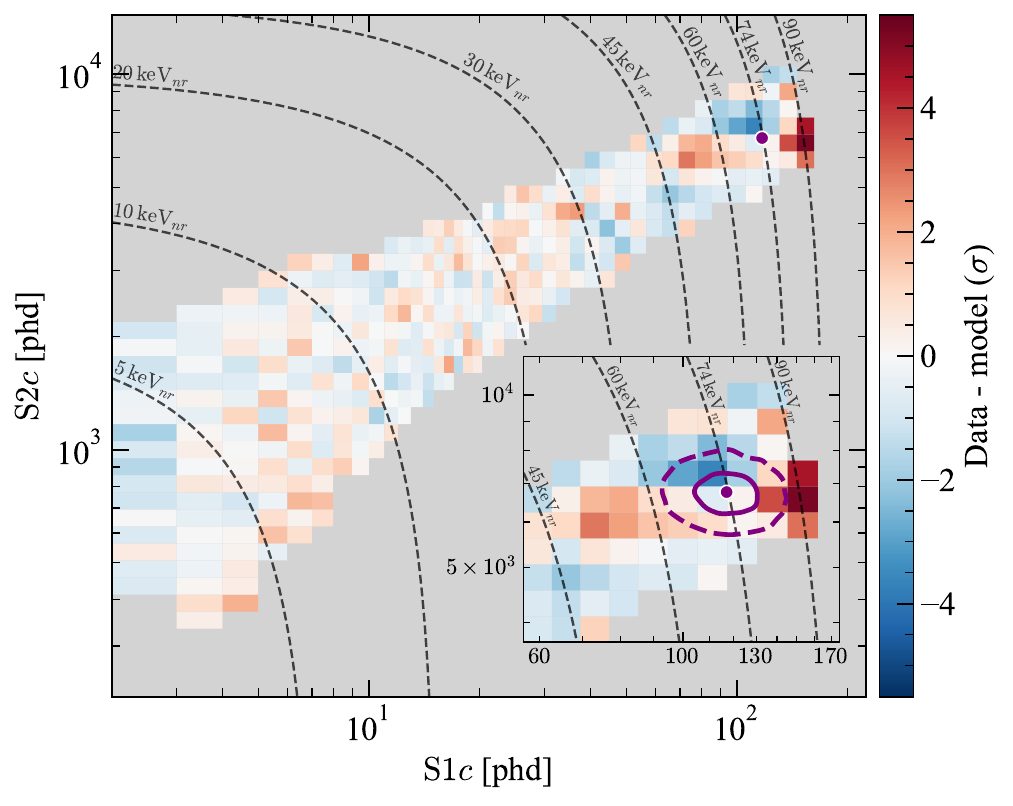}
  \hfill
    \includegraphics[width=0.49\textwidth]{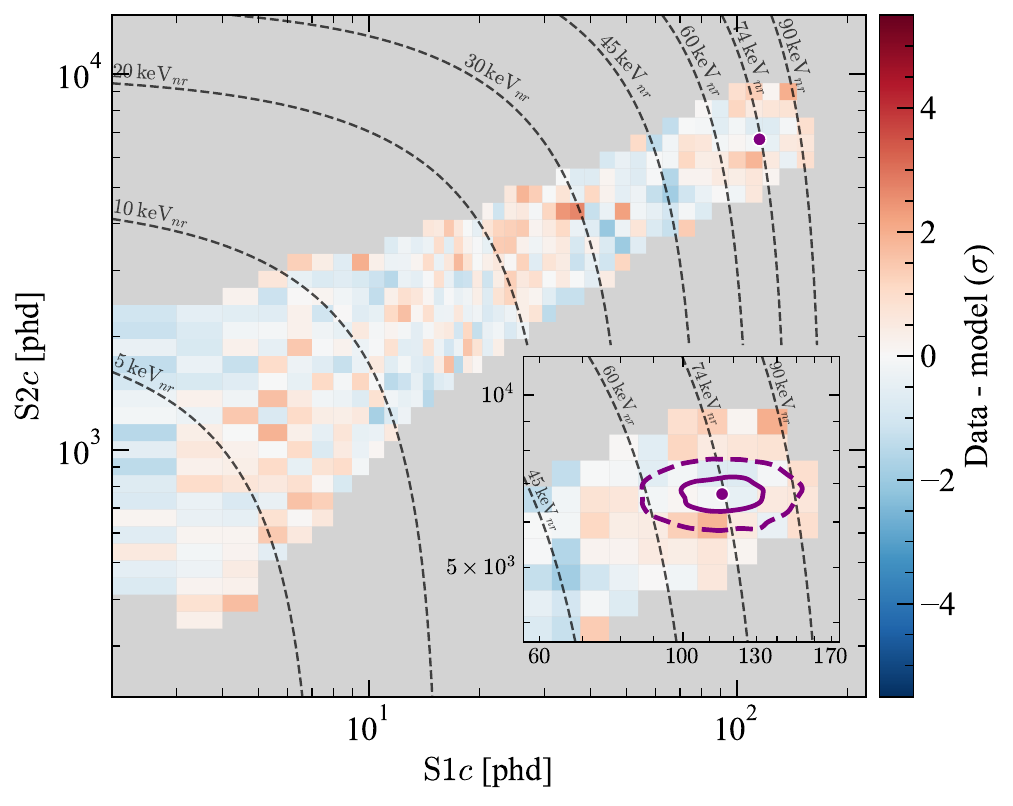}
    \label{fig:2Dgof_bestfit}
  \caption{Two-dimensional goodness-of-fit in the \(\log_{10}(\mathrm{S1}c)\)-\(\log_{10}(\mathrm{S2}c)\) plane for DD calibration data, shown as bin-by-bin signed residuals in units of equivalent Gaussian significance. Only bins with either the data count or the model median count greater than or equal to five are shown. Red represents a data excess and blue represents a data deficit. In each panel, the purple point shows the model-predicted mean \((\mathrm{S1}c,\,\mathrm{S2}c)\) for nuclear recoils at the \(74~\mathrm{keV}\) endpoint. The purple contours in the insets are density contours of simulated single-scatter \(74~\mathrm{keV}\) events generated from the corresponding model after applying the analysis selections and efficiencies. The enclosed probability levels are 0.393 and 0.865, corresponding to the $1\sigma$ and $2\sigma$ regions of a two-dimensional Gaussian. \textbf{Left:} Model with fluctuation parameters of default \textsc{nest}; where $F_\mathrm{ex}=0.4$, $F_\mathrm{i}=0.4$, $\lambda=1$, $A=0.04$ are held fixed and all other parameters are sampled. Localized data excesses appear in the inset bins to the left ($3.0\sigma$) and right ($5.2\sigma$) of the DD endpoint, respectively, while a localized data deficit ($3.6\sigma$) appears above the endpoint. \textbf{Right:} Best-fit model with updated fluctuation modeling and all parameters floated.} 
  \label{fig:2DGoF}
\end{figure*}

Figure~\ref{fig:2DGoF} compares the DD data to models with and without the proposed fluctuation modifications in two-dimensional (S1$c$, S2$c$) space. 
While a \textsc{nest}-like model ($F_\mathrm{ex}=0.4$, $F_\mathrm{i}=0.4$, $\lambda=1$, $A=0.04$) describes data with energies below the DD endpoint well, it exhibits localized disagreements near the endpoint.
Our modified treatment of fluctuations resolves this discrepancy. Specifically, the enhanced $F_\mathrm{ex}$ increases the S1$c$ variance, while the sub-binomial recombination narrows the S2$c$ distribution. 
The model successfully describes the location and width of the data distribution over two orders of magnitude in event counts, from the high-statistics endpoint to the low-statistics threshold region in the full 2–74~keV range. 

We verify that both modifications are required to resolve the disagreement at the endpoint. 
When fixing recombination variance in \textsc{nest} to $\lambda=1$ and $A=0.04$ while floating all other parameters, the fit attempts to compensate by suppressing primary ion fluctuations ($F_\mathrm{i}$), but still yields a projected S2$c$ reduced-$\chi^2$ of $1.7$ compared to $0.8$ for the best-fit model.
Similarly, removing the energy dependence from exciton fluctuations and holding $F_\mathrm{ex}=0.4$ prevents the model from describing the S1$c$ spectrum above 60~phd, resulting in a projected S1$c$ reduced-$\chi^2$ of $3.9$, compared to the best-fit model's $1.7$.

\begin{figure*}
  \centering
    \includegraphics[width=0.7\textwidth, trim={1cm 1cm 1cm 1cm}, clip]{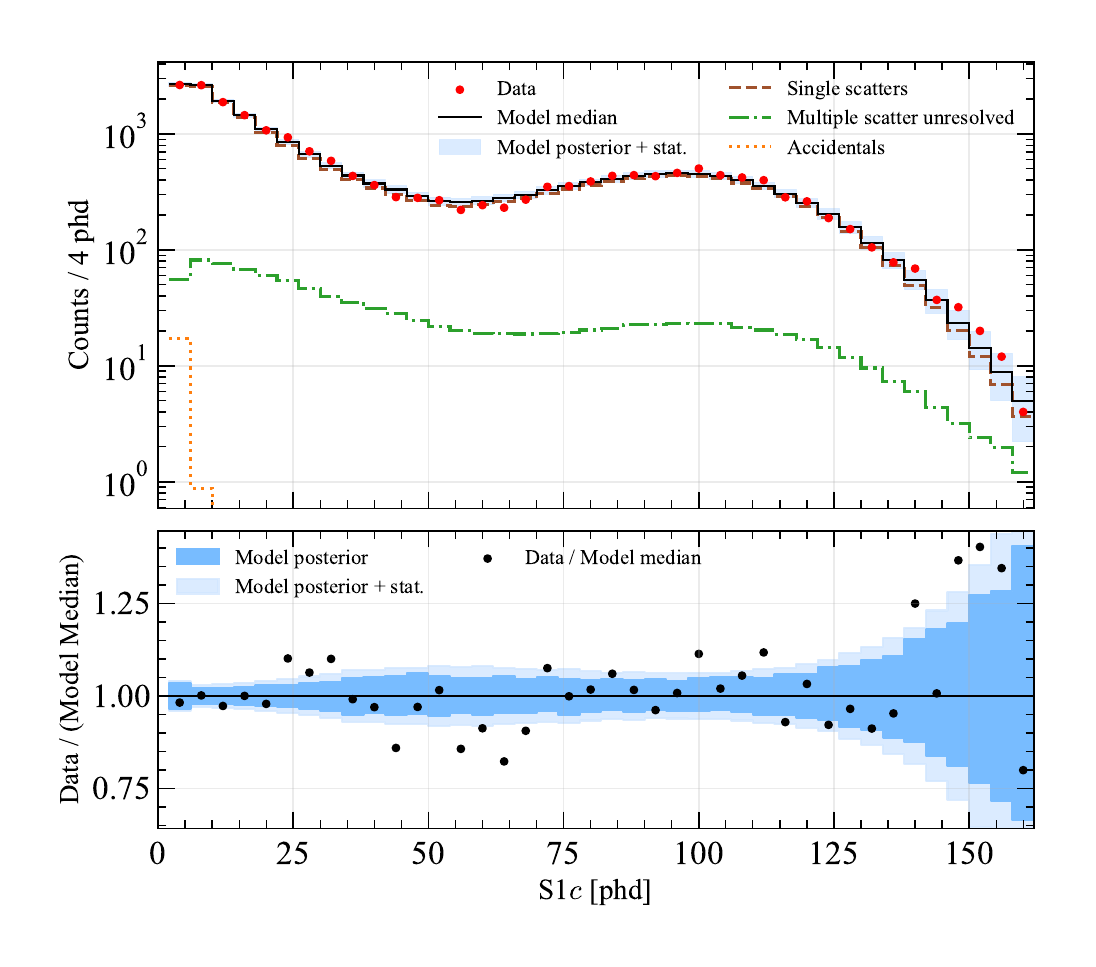}
    \\
    \vspace{0.1cm}
    \includegraphics[width=0.7\textwidth, trim={1cm 1cm 1cm 2cm}, clip]{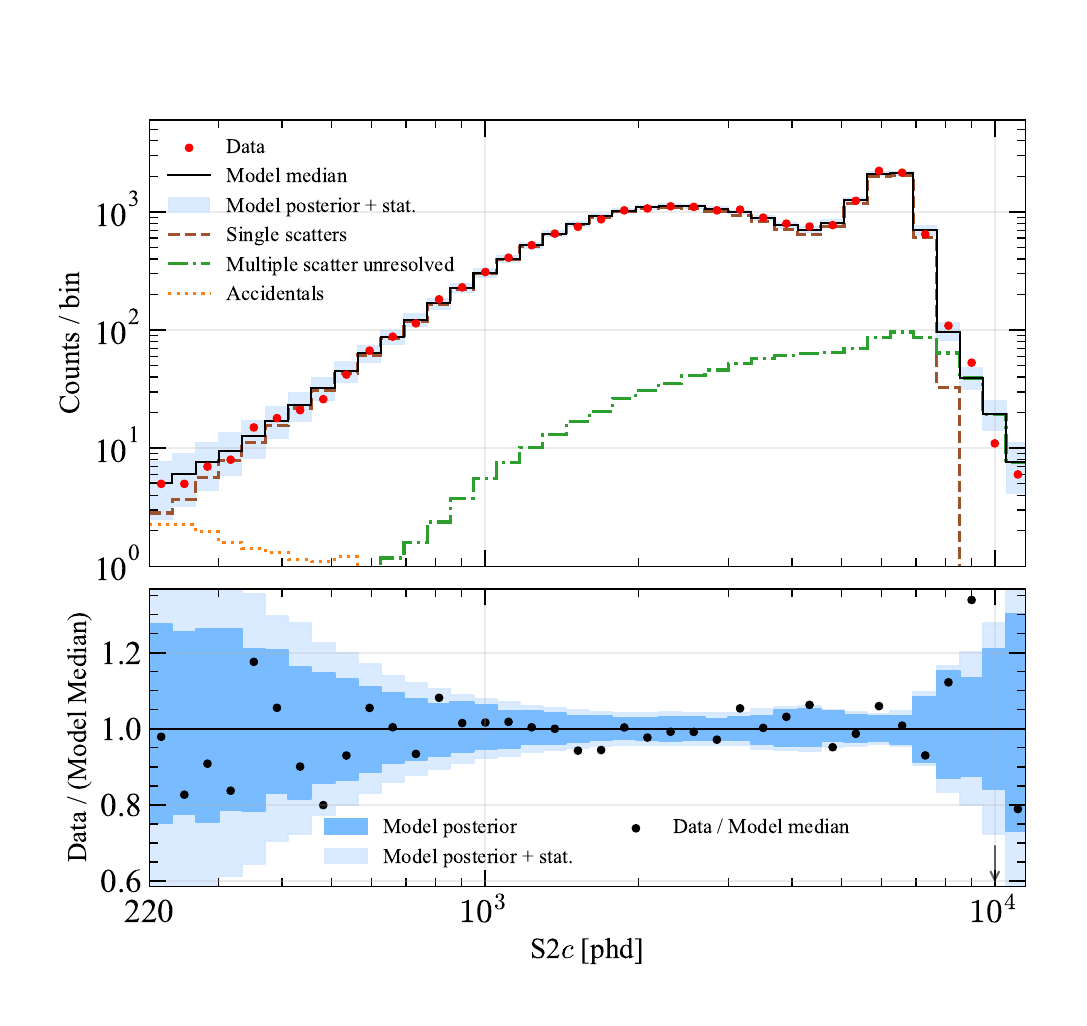}
  \caption{Projections in S1$c$ (top) and S2$c$ (bottom) of the data as the red points in the full ROI and comparison with the best-fit model as the solid black lines. The dashed brown lines denote the SS component and the dot-dashed green lines denote the UMS component. The UMS contribution is necessary to account for the high-S2$c$ tail in the data. The orange dotted lines denote the accidentals component which become prominent at low S1$c$ and S2$c$. The final model prediction is the sum of the SS, UMS, and accidental components. 
  The dark (light) blue shaded bands indicate the posterior model (posterior model plus statistical) uncertainties, which are larger in regions of lower counts.
  The black points in the lower panels show the ratio of data to model.
  }
  \label{fig:gof_full}
\end{figure*}

\begin{figure*}
  \centering
    \includegraphics[width=0.7\textwidth, trim={1cm 1cm 1cm 1cm}, clip]{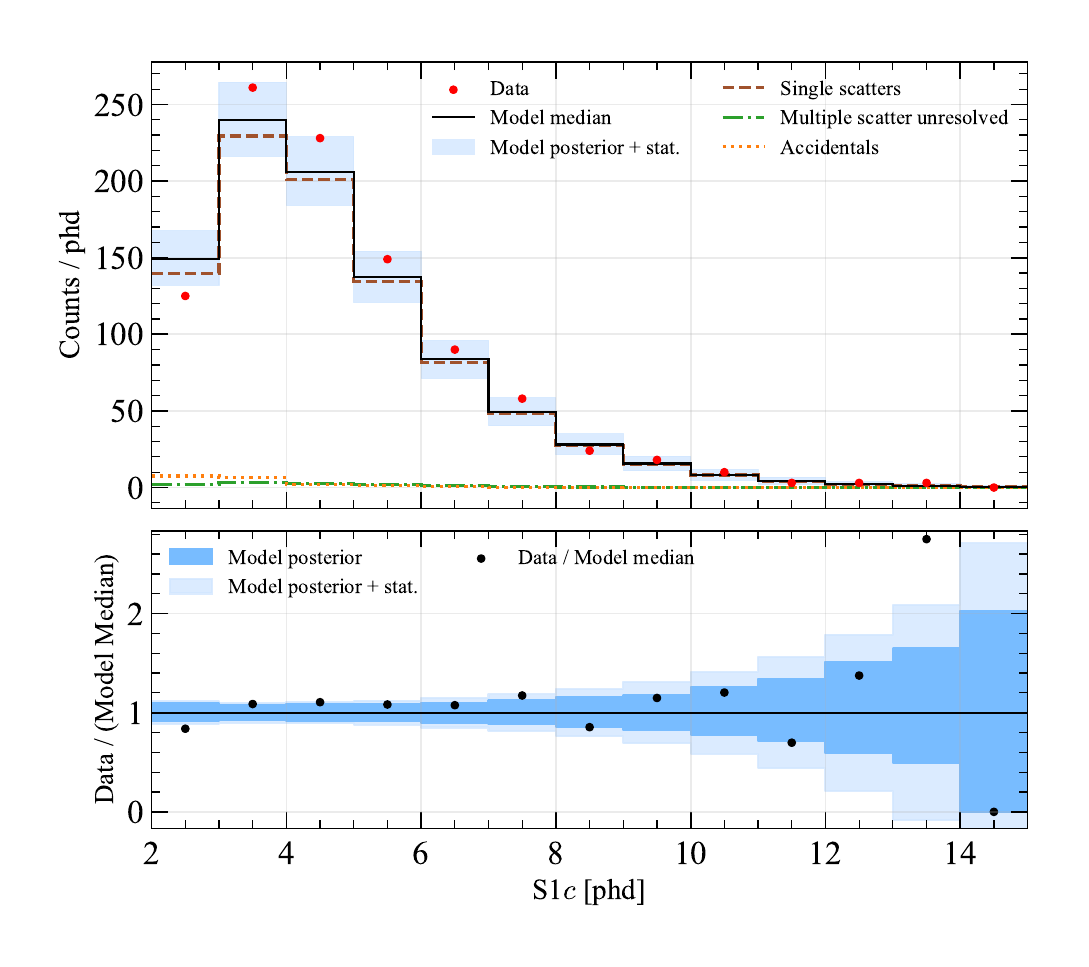}
  \\
  \vspace{0.1cm}
    \includegraphics[width=0.7\textwidth, trim = {1cm 1cm 1cm 2cm}, clip]{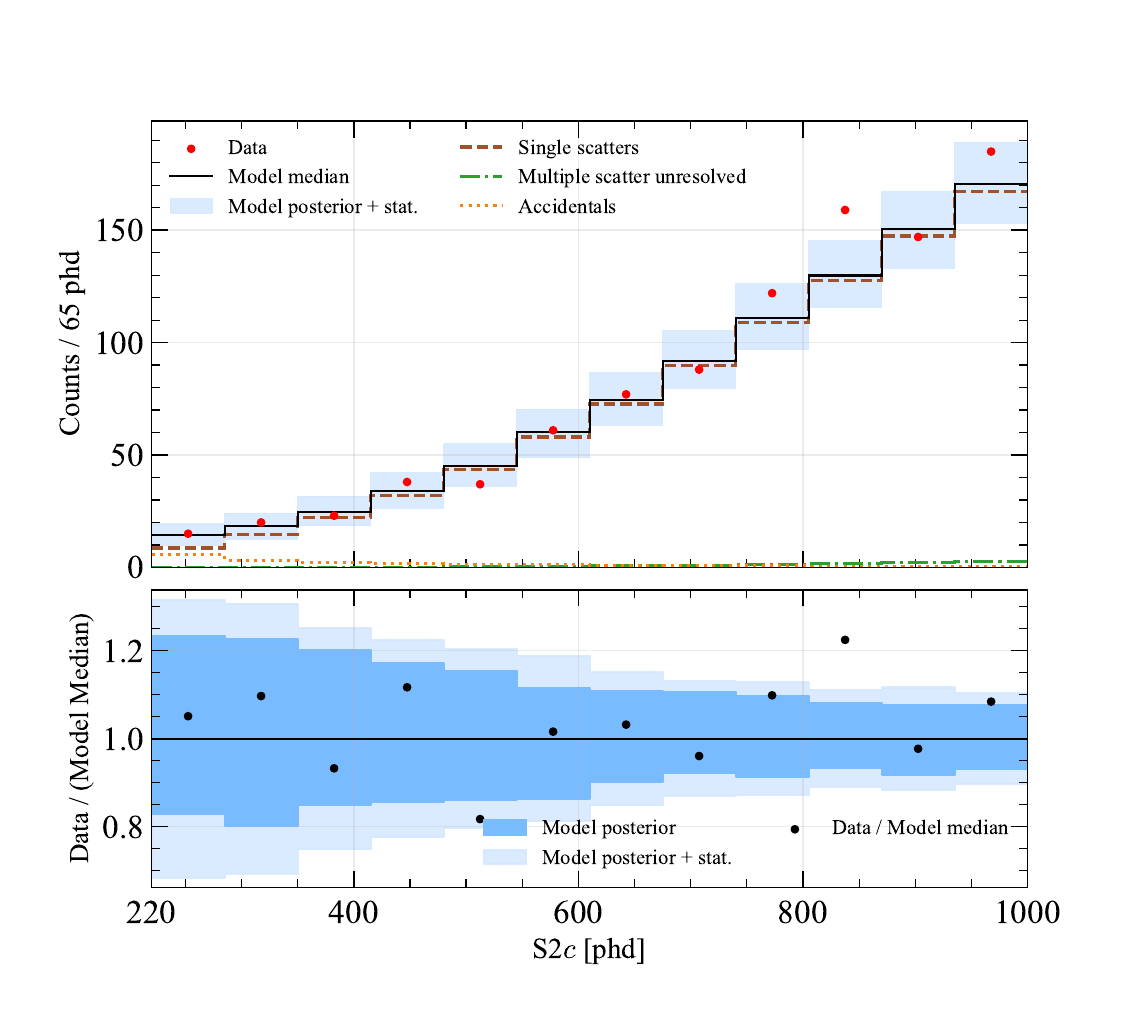}
  \caption{Projections in S1$c$ (top) and S2$c$ (bottom) of the data as the red points in the restricted ROI compatible with the low-energy analysis from Ref.~\cite{akerib2025searches} and comparison with the best-fit model as the solid black lines. The dashed brown lines denote the SS component, the dot-dashed green lines denote the UMS component, and the orange dotted lines denote the accidentals component.
  The final model prediction is the sum of the SS, UMS, and accidental components. The dark (light) blue shaded bands indicate the posterior model (posterior model plus statistical) uncertainties, which are larger in regions of lower counts. The black points in the lower panels show the ratio of data to model.  
  We calculate posterior predictive p-values~\cite{gelman1996posterior_posteriorPredictiveModel} in this region and obtain $0.15$ and $0.34$ for projections in S1$c$ and S2$c$, respectively.
  }
  \label{fig:gof_lowe}
\end{figure*}

Comparisons of data and best-fit model in projected S1$c$ and S2$c$ spaces are shown throughout the entire energy range in Fig.~\ref{fig:gof_full} and in a low-energy region of interest relevant to the LZ low-mass WIMP and $^8$B result \cite{akerib2025searches} in Fig.~\ref{fig:gof_lowe}.
The model uncertainties in these comparisons are obtained by sampling the MCMC posterior, including both \textsc{nest} and neutron simulation parameters.

The posterior value for the sub-2~MeV fraction of $0.11 \pm 0.03$ is consistent with a dedicated simulation and an independent characterization of the DD generator, both reported in Ref.~\cite{CaseyRhyneThesis}.
Likewise, the posterior UMS fraction of $(5.0^{+1.2}_{-1.1})\%$ is consistent with simulation and a data-driven estimate in LZ.
The neutron-Xe elastic differential cross section is modeled as a linear combination of the JENDL-4.0u and ENDF/B-VII.1 nuclear libraries shown in Fig.~\ref{fig:recoil_spectrum_g4ndl_jendl} to provide a reasonable range of prior uncertainty. 
The MCMC samples the single mixture parameter in this linear ansatz.
We find in the posterior a mixture with $(63^{+11}_{-16})\%$ of the JENDL-4.0u library.
As the shape of the cross section is constrained only through the single degree of freedom in the linear mixture, we do not interpret the posterior as a true constraint on the differential cross section.
Rather, the mixture fraction acts as a nuisance parameter that propagates the difference between the two libraries into the uncertainties on the yield and fluctuation parameters.
This parameterization does not capture variations in the cross-section shape outside the span of the two libraries.
We note that the endpoint energy is fixed by the kinematics alone, so the cross-section does not affect the energy scale but only how the events are distributed below the endpoint.
The $30-50$~keV region of the spectrum is the most impacted by the choice of nuclear library, and ultimately leads to the rejection of pure-ENDF/B-VII.1 and pure-JENDL-4.0u samples. 

\begin{figure}
    \centering
    \includegraphics[width=0.95\linewidth]{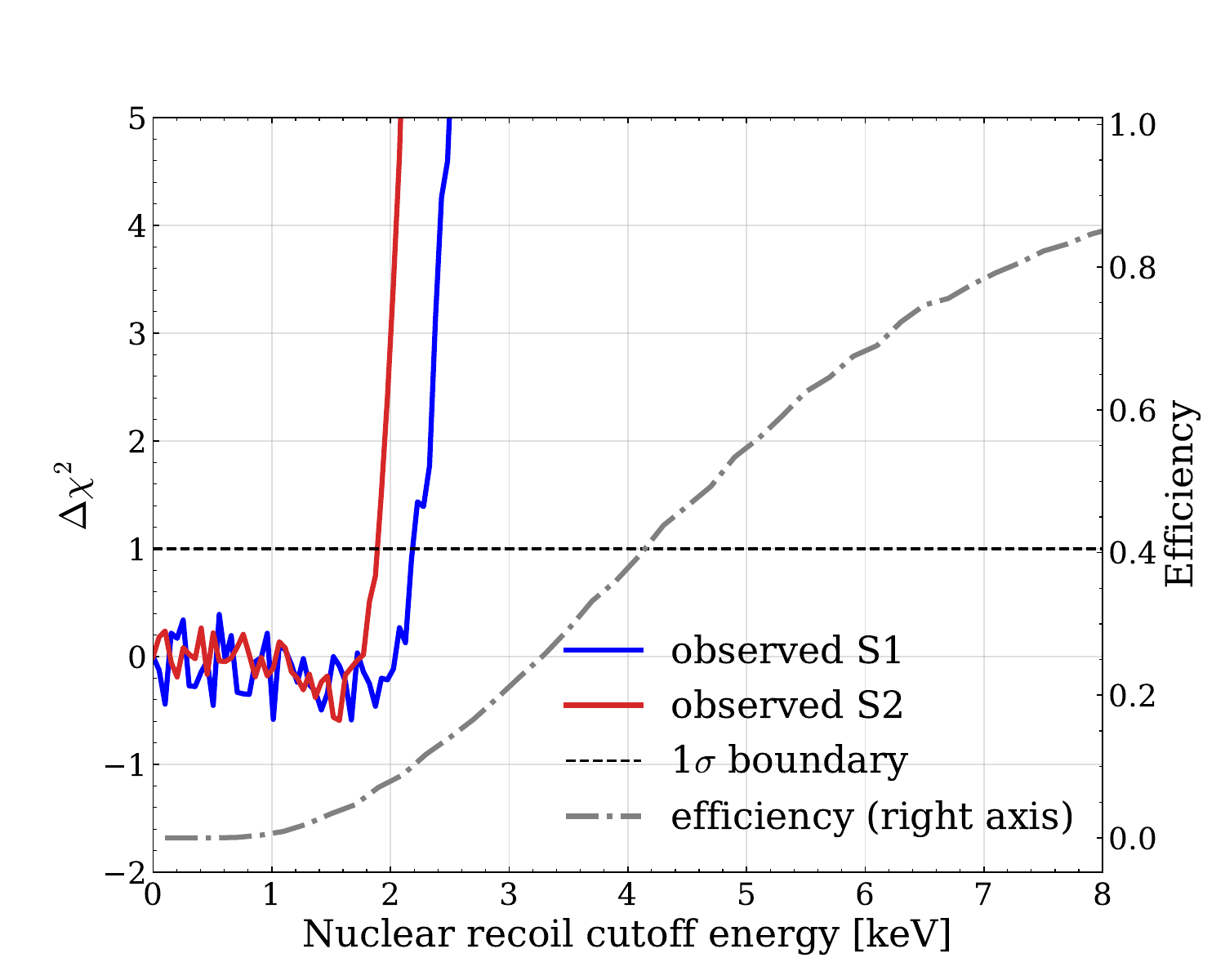}
    \caption{$\Delta \chi^{2}$ for 1D projections of the data against models with a cutoff in yields increasing with energy. The y-axis corresponds to a $1$-$\sigma$ fluctuation of the base model, as computed with Monte Carlo toys, to account for possible differences in degrees of freedom for each projection. The $\Delta \chi^{2}$ reaches a $1$-$\sigma$ discrepancy from the base model at an energy cutoff of 2.2 keV and 1.9 keV for the S1 and S2 projections, respectively, giving an estimate of the energy threshold for sensitivity to yields from each channel. The dot-dashed gray line, to be read with the secondary y-axis on the right, illustrates the corresponding efficiency in selecting events due to reconstruction at a given energy, driven primarily by the threefold coincidence requirement for the S1 and estimated through detector response simulations. The low value of efficiency at small energies is responsible for the loss in sensitivity.
    }
    \label{fig:energy_sens}
\end{figure}

\begin{figure}
    \centering
    \includegraphics[width=0.95\linewidth]{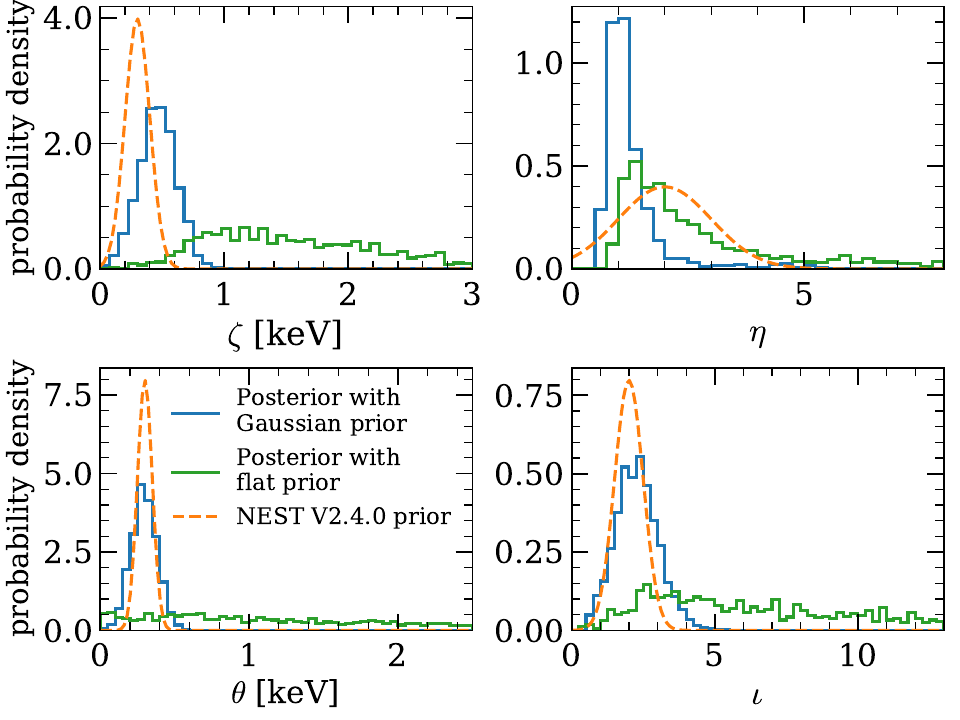}
    \caption{Gaussian priors and marginalized posteriors for the four low energy yield parameters in the model. The top and bottom rows regard parameters affecting charge and light yield, respectively. Posteriors in blue are those used in this result, and are constrained by Gaussian \textsc{nest} priors. Posteriors in green illustrate how, in the case of a flat prior, the model uncertainty is significantly higher.}
    \label{fig:fantastic_four_prior}
\end{figure}

The energy sensitivity threshold for this analysis was estimated by computing the $\chi^{2}$ goodness-of-fit of projections of the data in S1$c$ and S2$c$ space against a modified version of the model in which yields were set to zero below a variable energy threshold. The constraining power for yields below these energies is limited due to the 3-fold selection. The results of these threshold tests are shown in Fig.~\ref{fig:energy_sens}. 
The threshold at which the modified model first shows a clear discrepancy with the data gives a low-energy sensitivity of 2.2~keV in S1$c$ and 1.9~keV in S2$c$.
Appendix~\ref{app:threshold} discusses the data and model near the threshold in more detail. 
Figure~\ref{fig:fantastic_four_prior} shows the priors for $\theta$, $\iota$, $\zeta$ and $\eta$, and shows how the posteriors are still influenced by the data in this energy regime despite lower statistics. 

%% file: Discussion.tex
\section{Discussion}
\label{sec:discussion}

We now highlight the main contributions of detector and LXe response parameters to the S1$c$ and S2$c$ variance. 
The variance contributions are quoted in percentage of squared resolutions \(\operatorname{Var}(\text{S1}c)/\langle \text{S1}c\rangle^{2}\) and \(\operatorname{Var}(\text{S2}c) / \langle \text{S2}c \rangle^{2}\). 
The default \textsc{nest~v2.4.0} model's primary source of S1$c$ variance for NR events is from binomial photon detection fluctuations, constituting 78\% of the total squared resolution across all energies, with $F_\text{ex}=0.4$ contributing 2--3\% across all energies. 
The remaining fluctuations come from the single photon response of the PMTs, ionization fluctuations, and recombination fluctuations. 
In our modified NR model, the binomial photon detection contribution is 73\% and $F_\text{ex}$ is 10\% at 3~keV.
We see the biggest impact of the energy-dependent $F_\text{ex}$ at the DD endpoint, which makes the total squared resolution for S1$c$ in our modified model to be 1.6 times that of the default \textsc{nest~v2.4.0} model.
The relative contributions of the photon detection and $F_\text{ex}$ evolve to 47\% and 41\% in our modified model at the DD endpoint, respectively.
Recombination fluctuations, regardless of whether supra-binomial (similar to \textsc{nest~v2.4.0}) or sub-binomial as suggested in this work, are always subdominant ($\leq 2$\%) across all energies for S1$c$ resolution.

The S2$c$ variance is governed by ionization fluctuations (via $F_\text{i}$), recombination fluctuations, and the electron extraction, with electroluminescence fluctuations and the single photon response of the PMTs contributing at subdominant ($<5$\%) levels.
In our modified model the total squared resolution at the endpoint is 0.73 times the value predicted by default \textsc{nest~v2.4.0}.
We suggest that sub-binomial recombination fluctuations are responsible for the lower S2$c$ variance, since fluctuations that are binomial or larger fail to describe the endpoint as shown in Fig.~\ref{fig:2DGoF}. 
We find that binomial recombination fluctuations may be recovered if associated with zero ionization fluctuations at the endpoint.
An electron extraction efficiency above 91\% could explain the low observed S2$c$ resolution, but this efficiency is inconsistent with the 73\% measured in data.

The variances of ER events verify that the fluctuation response presented in this work is specific to NR events, and do not arise from common detector effects.
Specifically, we checked ER models used in previous LZ WIMP searches~\cite{PhysRevLett.131.041002_WS2022Results, 4dyc-z8zf_WS2024Results} (tuned to calibrations using \textsc{nest~v2.4.0}~\cite{szydagis_2023_8215927_NEST_2-4-0}), data from injected tritium $\beta$ decays, the ${}^{131\text{m}}$Xe conversion electron at 164~keV, and the K, L, and M (S2-only) atomic de-excitations of ${}^{125}$Xe.
Enhanced S1$c$ fluctuations and suppressed S2$c$ fluctuations are not required in these cases. 
We investigated the position selection as another factor that impacts the recoil energy spectrum, and verified with a separate MCMC that it does not negate the sub-binomial recombination and energy-dependent $F_\text{ex}$. 
We further verified that these modifications are still required when using the pure-ENDF/B-VII.1 and pure-JENDL-4.0u nuclear libraries. 
Finally, we stress-tested the MCMC framework with fits against simulated data, and recovered posteriors that were consistent with the parameter sets used in the simulations. 

A DD run conducted in 2021 when LZ was operating at a drift field of 193~V/cm, and another conducted in 2023 at 97~V/cm, both exhibit the enhanced S1 variance and suppressed S2 variance seen in this work.
Although the LZ YBe analysis~\cite{aalbers2025lowenergynuclearrecoilcalibration_LZYBe} also focused on low-energy recoils, the dataset does not have sufficient statistical power to constrain \(F_{\mathrm{ex}}\).
There, increasing \(F_{\mathrm{ex}}\) from 0.4 to 3 changes the predicted YBe rate by only 9\%, which is subdominant to the 23\% total uncertainty reported in Ref.~\cite{aalbers2025lowenergynuclearrecoilcalibration_LZYBe}.
Since the YBe recoil energy range is limited to be below 5~keV, this calibration is not sensitive to the shape of \(F_{\mathrm{ex}}\). 

Having established that the DD calibration data require modified treatments of exciton and recombination fluctuations, we evaluate the implications for solar neutrino searches.
In the DD-tuned model, the total prediction of $19.6^{+8.5}_{-6.5}$ ${}^8$B events is higher than the \textsc{nest~v2.4.0} default ($11.6^{+8.4}_{-5.7}$). Of this increase, about four events arise from the higher $L_{\mathrm{y}}$ inferred from DD data; the enhanced $F_{\mathrm{ex}}$ contributes four additional $^8$B events compared to a model with $F_{\mathrm{ex}}=0.4$. Although the DD-inferred $L_{\mathrm{y}}$ is only 13\%--37\% higher at the relevant recoil energies, the resulting increase in the expected ${}^8$B count is much larger because these low-energy events lie close to threshold and rely on upward S1 fluctuations to satisfy the three-fold coincidence requirement; a modest increase in the mean S1 therefore produces a disproportionately large increase in the acceptance.
Other fluctuation parameters, including $F_{\mathrm{i}}$ and $\lambda$, have a negligible impact on the expected ${}^8$B event rate. 
The model used in Ref.~\cite{akerib2025searches} predicted $20.6^{+8.9}_{-6.8}$ events; the updated calibration model presented here lowers this by about one event. 
This comes from two changes to the calibration model: the DD accidentals fraction is now floated instead of fixed at its upper limit, and the DD event position sampling has been corrected. 
The resulting shift is small compared to the NR model uncertainty on the predicted $^8$B rate and does not affect the conclusions of Ref.~\cite{akerib2025searches}.

The modifications to the LXe response model have implications in searches for WIMPs above 10~GeV as well. 
Enhanced exciton-production fluctuations (\(F_{\mathrm{ex}}\)) and suppressed recombination variance (\(\lambda\)) change the predicted two-dimensional \((\mathrm{S1}c,\mathrm{S2}c)\) distribution for NR signals, particularly the behavior of WIMP tails.
This matters for analyses using the profile-likelihood-ratio, which is sensitive to the shape of the signal model in \((\mathrm{S1}c,\mathrm{S2}c)\) space. 
Accurately modeling the signal shape is important to minimize bias in signal acceptance and background discrimination.

A first-principles explanation for the observed exciton and recombination fluctuations is beyond the scope of this work. 
However, the DD calibration data provide clear constraints that can be tested with future measurements. 
In particular, low-energy neutron calibrations such as D-reflector configurations~\cite{CaseyRhyneThesis, Aalbers_2024_DesignImplementationPerformanceLZCalibration}, and analyses employing reduced PMT coincidence thresholds will enable further tests of the energy scaling and field dependence of the fluctuation model introduced here.

%% file: Conclusion.tex
\section{Conclusion}

The LXe NR response with enhanced fluctuations in pre-recombination exciton production and suppressed recombination fluctuations was calibrated using a 2.45 MeV DD neutron source in the LZ experiment. 
Validation checks were conducted and indicate that these modifications are required to fit the DD data. This enhanced pre-recombination exciton fluctuation allows the lowest-energy events to more frequently meet threshold requirements for detection without significantly impacting the signal shape. 
The combination of these two fluctuation modifications affects the signal shape of high-energy nuclear recoils without affecting the signal rate. 
Therefore, accurate treatment of these fluctuations is essential for interpreting future NR searches in LXe experiments.
This DD calibration of mean yields and fluctuation was used in the modeling and searches of 3-9 GeV/c${}^{2}$ low-mass WIMP and $^8$B CE$\nu$NS in LZ~\cite{akerib2025searches}.

%% file: Acknowledgements.tex
\begin{acknowledgments}
The research supporting this work took place in part at the Sanford Underground Research Facility (SURF) in Lead, South Dakota. Funding for this work is supported by the U.S. Department of Energy, Office of Science, Office of High Energy Physics under Contract Numbers DE-AC02-05CH11231, DE-AC02-76SF00515, DE-AC52-07NA27344, DE-SC0008475, DE-SC0010004, DE-SC0010010, DE-SC0010072, DE-SC0011702, DE-SC0012447, DE-SC0012704, DE-SC0014223, DE-SC0015535, DE-SC0015910, DE-SC0018982, DE-SC0019066, DE-SC0019193, DE-SC0020216, DE-SC0025446, DE-SC0025629, DE-SC0026014, DE-SC0026544. This research was also supported by U.S. National Science Foundation (NSF) ; the UKRI’s Science \& Technology Facilities Council under award numbers ST/M003639/1, ST/R003181/1, ST/W000466/1, ST/W000490/1, ST/W000555/1, ST/W00058X/1, ST/W000628/1, ST/W000636/1, ST/X00600X/1, ST/X006050/1, UKRI2837, UKRI2840, UKRI2841, UKRI2842, UKRI2843, UKRI2844, UKRI2847, UKRI2851, UKRI2852; the Portuguese Foundation for Science and Technology (FCT) under award number PTDC/FIS-PAR/2831/2020; the Institute for Basic Science, Korea (budget number IBS-R016-D1); the Swiss National Science Foundation (SNSF) under award number 10001549; the European Union’s research and innovation programme Horizon Europe under the Marie Skłodowska-Curie grant agreement No 101209429. This research was supported by the Australian Government through the Australian Research Council Centre of Excellence for Dark Matter Particle Physics under award number CE200100008. We acknowledge additional support from the UK Science \& Technology Facilities Council (STFC) for PhD studentships and the STFC Boulby Underground Laboratory in the U.K., the GridPP~\cite{faulkner2005gridpp, britton2009gridpp} and IRIS Collaborations, in particular at Imperial College London and additional support by the University College London (UCL) Cosmoparticle Initiative, and the University of Zurich. We acknowledge additional support from the Center for the Fundamental Physics of the Universe, Brown University. K.T. Lesko acknowledges the support of the Royal Society, the Wolfson Foundation, and Brasenose College. This research used resources of the National Energy Research Scientific Computing Center, a DOE Office of Science User Facility supported by the Office of Science of the U.S. Department of Energy under Contract No. DE-AC02-05CH11231. We gratefully acknowledge support from GitLab through its GitLab for Education Program. Fermilab is managed by Fermi Forward Discovery Group, LLC, acting under Contract No. 89243024CSC000002. The University of Edinburgh is a charitable body, registered in Scotland, with the registration number SC005336. The assistance of SURF and its personnel in providing physical access and general logistical and technical support is acknowledged. We acknowledge the South Dakota Governor's office, the South Dakota Community Foundation, the South Dakota State University Foundation, and the University of South Dakota Foundation for use of xenon. We also acknowledge the University of Alabama for providing xenon. For the purpose of open access, the authors have applied a Creative Commons Attribution (CC BY) license to any Author Accepted Manuscript version arising from this submission. Finally, we respectfully acknowledge that we are on the traditional land of Indigenous American peoples and honor their rich cultural heritage and enduring contributions. Their deep connection to this land and their resilience and wisdom continue to inspire and enrich our community. We commit to learning from and supporting their effort as original stewards of this land and to preserve their cultures and rights for a more inclusive and sustainable future. 
                                                                                     
\end{acknowledgments}

%% file: Appendix.tex
\section*{Appendix}
\subsection{Accidentals rate modeling}\label{sec:accidentals}
In the LZ low-mass WIMP (LMW) and \(^8\)B neutrino searches~\cite{akerib2025searches}, the DD calibration model fixed the \emph{DD accidentals} component to the upper-bound driven by data of 3.7\% in \(2<\mathrm{S1}c<10~\mathrm{phd}\) and \(220<\mathrm{S2}c<1000~\mathrm{phd}\). In the present work, this DD accidentals normalization is instead floated in the MCMC, resulting in a best-fit fraction of \(1.9^{+1.2}_{-1.3}\%\). We emphasize that this DD accidentals fraction is a calibration nuisance parameter and is distinct from the accidental-background treatment in the LMW and \(^8\)B searches themselves. We also emphasize that the signal sensitivity and \((\mathrm{S1}c,\mathrm{S2}c)\) signal-shape in those analyses are insensitive to whether the DD accidentals normalization is fixed or floated within this range. 

The model used in the $^8$B search predicted a rate that is $\sim5$\% higher than the model presented in this work, which has both the update to the accidentals rate and a correction in the position sampling. 
This shift is small and within the uncertainty quoted from the NR-model on the expected \(^8\)B rate \((-33\%,+43\%)\). 
For the \(^8\)B search, this choice primarily affects the \emph{expected} sensitivity (or expected significance) through the predicted signal rate, while the \emph{observed} significance is driven by the data and the fitted signal-plus-background model and is therefore not materially altered by this calibration convention. 
A change of 5\% in the $^8$B rate does not significantly impact the result in Ref.~\cite{akerib2025searches}. 

\subsection{Correlation matrix}
\label{app:corr}

\begin{figure*}[t]
    \centering
    \includegraphics[width=0.9\textwidth]{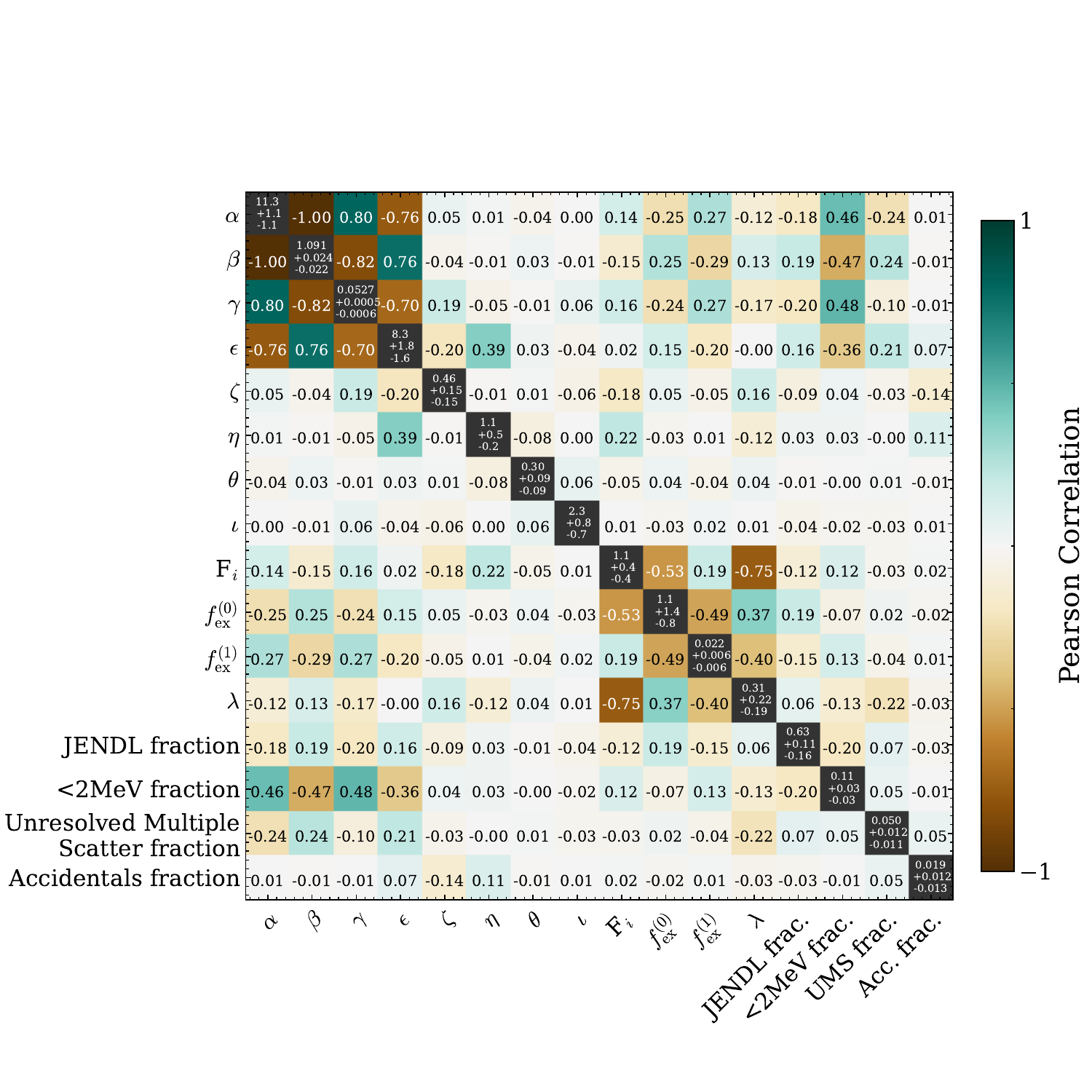}
    \caption{Summary of the best fit to the data. The diagonal entries show the median of the posterior for each parameter together with the corresponding 1-\(\sigma\) credible interval. The off-diagonal entries show the Pearson correlation coefficient for each parameter pair, with the cell background color providing a visual representation of the same quantity.}
    \label{fig:best-fit-parameters}
\end{figure*}

Figure~\ref{fig:best-fit-parameters} summarizes the best fit to the data. The diagonal entries show the median of the posterior for each parameter together with the corresponding 1-\(\sigma\) credible interval. The off-diagonal entries show the Pearson correlation coefficient for each parameter pair, with the cell background color providing a visual representation of the same quantity. Strong correlations and anticorrelations are present among the mean-yield parameters \(\alpha\), \(\beta\), \(\gamma\), and \(\epsilon\). The anticorrelation between \(F_\mathrm{i}\) and \(\lambda\) is expected because both affect the \(\mathrm{S2}\) variance, with \(F_\mathrm{i}\) playing the larger role at lower nuclear-recoil energies and \(\lambda\) becoming more important at higher energies, such as near the DD endpoint (\(74~\mathrm{keV}\)). The parameters \(F_{\mathrm{i}}\), \(f_{\mathrm{ex}}^{(0)}\), \(f_{\mathrm{ex}}^{(1)}\), and \(\lambda\) are also correlated because they all influence the width of the nuclear-recoil band. The model is able to disentangle the effects of \(\mathrm{S1}\) variance and \(\mathrm{S2}\) variance to some extent, although some degeneracy remains because both contribute to the band width.

\subsection{Reparametrization of uncertainty in yields}
\label{app:eigen}
\begin{figure}
    \centering
    \includegraphics[width=0.95\linewidth]{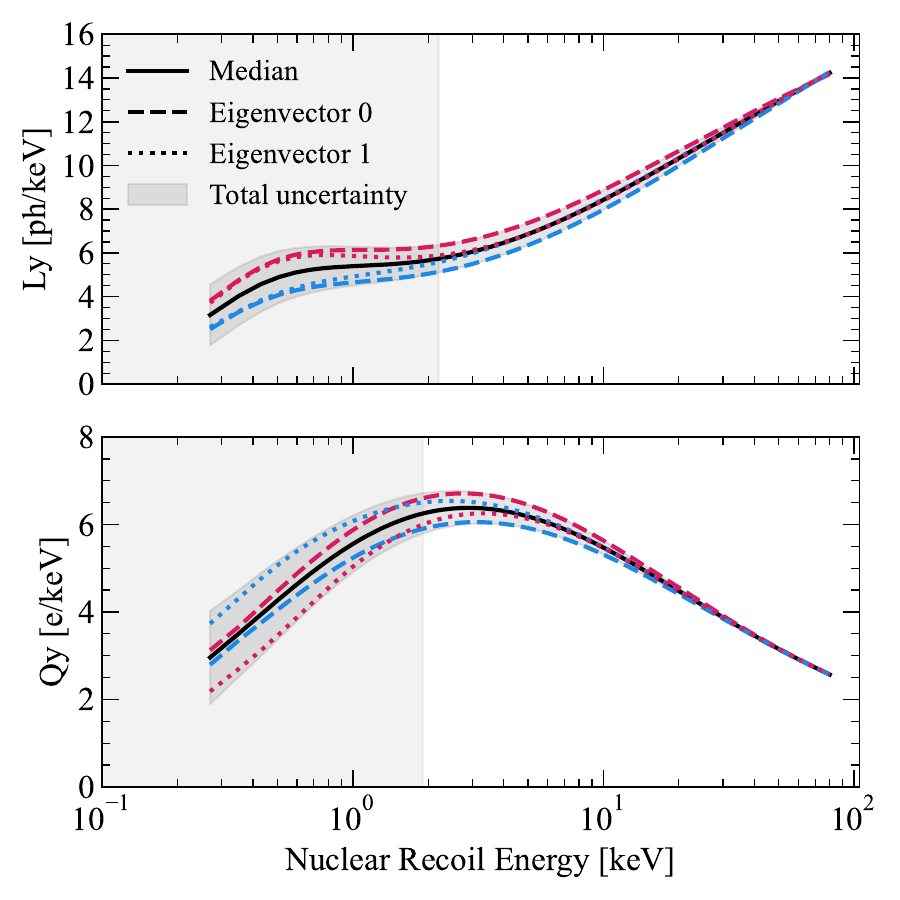}
    \caption{Median light and charge yields as computed from the MCMC chain along with uncertainties corresponding to the two most important eigenvectors scaled by their appropriate eigenvalues. The dark-gray bands are total uncertainty from all eigenvectors added in quadrature. The light-gray shaded area indicates the energy region where the data provides limited sensitivity to yields. The blue (red) line color represents the +1~(-1)~$\sigma$ excursions of the respective principal eigenvectors.}
    \label{fig:eigendecomposition}
\end{figure}

The number of free parameters and their correlations present a problem for any physics search based on such a model. In such an analysis, a minimal set of uncorrelated parameters with Gaussian uncertainties representing the variation in the yields is preferable. To this end, the following procedure was used to generate such a set. Independent samples of \textsc{nest} parameters were drawn directly from the MCMC chains, resulting in a collection of points in parameter space that followed the posterior distribution of the model. The corresponding light and charge yield for each set of parameters was computed at discrete energy values, resulting in an array of yields for each point in parameter space, equivalent to a set of yield curves consistent with the model posterior. The covariance matrix of the array of yields was computed, in effect treating the yield in each energy bin as if it were an independent (but possibly highly correlated) random variable. An eigendecomposition of the covariance matrix was performed (similar to Ref.~\cite{PhysRevD.97.012001_T2K_eigenDecompCitation}), with each eigenvector representing a distinct and orthogonal uncertainty mode for the collection of yields.

This procedure enables the parametrization of the uncertainty in light and charge yields simultaneously through a collection of eigenvectors in the combined light/charge yield space. The two most important eigenvectors were scaled by the appropriate eigenvalues to represent a $1\sigma$ fluctuation. These eigenvectors were handed off as the two main parameters controlling yields for the LMW/${}^{8}$B searches and are shown in Fig.~\ref{fig:eigendecomposition}. Eigenvectors 0 and 1 capture the principal correlated and anticorrelated components of the yield variance, respectively. The anticorrelated yield uncertainty increase in importance at lower energies for both charge and light yields below around 2~keV. 
Most of the uncertainty above this energy is carried by the correlated component.

\subsection{Threshold effects on the NR response}
\label{app:threshold}

\begin{figure}
    \centering
    \includegraphics[width=0.95\linewidth]{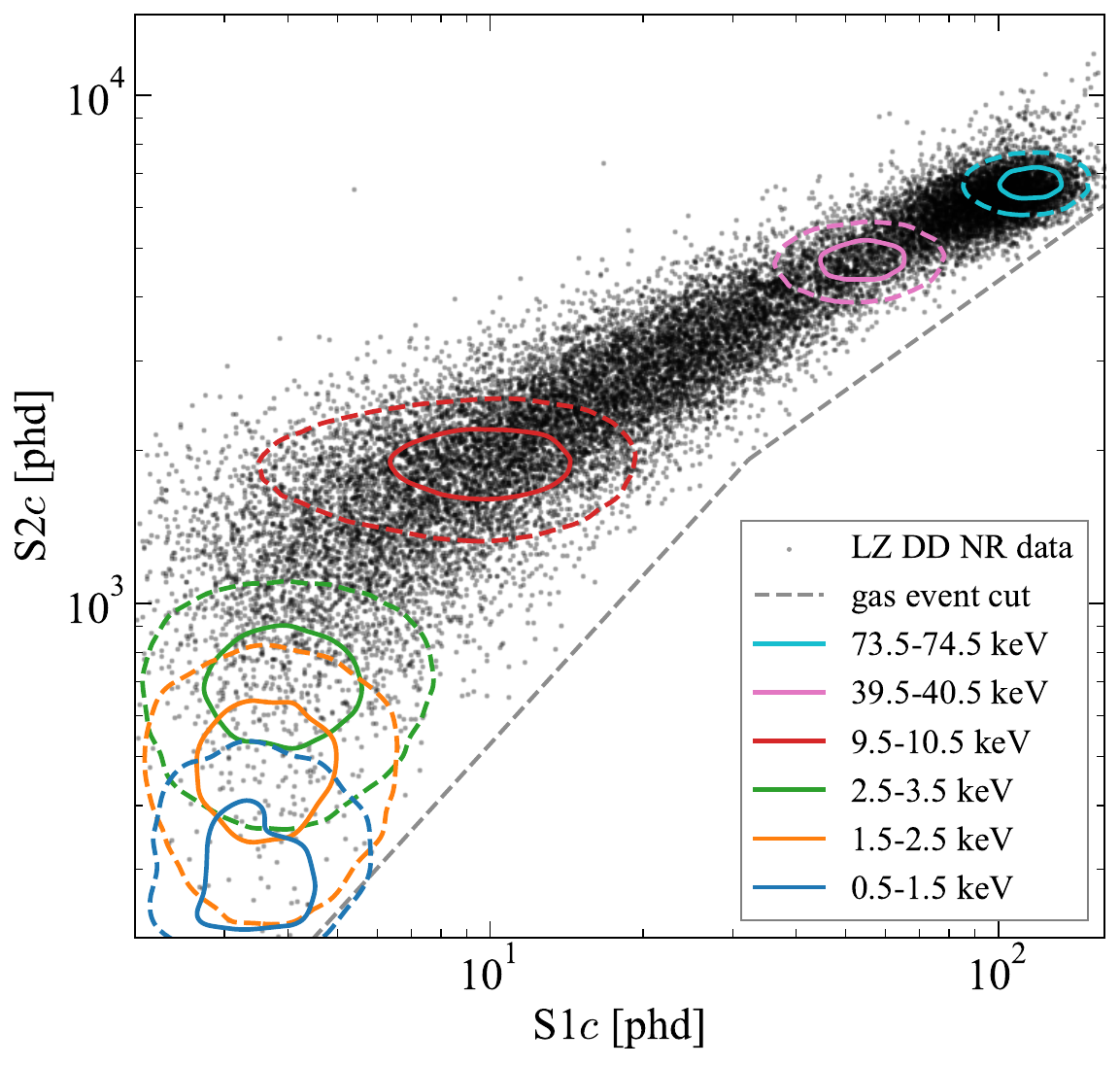}
    \caption{Final DD data (black points) in log--log $(\mathrm{S1}c,\mathrm{S2}c)$ space, overlaid with density contours for simulated single scatters from the best-fit NR response model after applying all efficiencies. Each color corresponds to a $1~\mathrm{keV}$-wide true recoil-energy interval. Solid and dashed contours enclose probabilities of 0.393 and 0.865, respectively, corresponding to the $1\sigma$ and $2\sigma$ regions of a two-dimensional Gaussian. The gray dashed line denotes the gas-event cut targeting n--GXe scatters. Table~\ref{tab:dd_contour_counts} gives the simulated component counts and DD data counts within each $1\sigma$ contour.}
    \label{fig:s1s2_contours}
\end{figure}

Detector and analysis efficiencies reshape the observed NR response near threshold. Figure~\ref{fig:s1s2_contours} illustrates this effect through density contours for simulated SS events after applying efficiencies, grouped into $1~\mathrm{keV}$-wide $E_{\mathrm{true}}$ slices. Table~\ref{tab:dd_contour_counts} gives the mean S1$c$ and S2$c$ signals and expected SS event counts both before and after applying efficiencies, together with the total SS efficiency for each $E_{\mathrm{true}}$ slice. It also compares the contributions from different model components with the observed DD data within each $1\sigma$ contour.

The lowest-energy $E_{\mathrm{true}}$ slices show the strongest changes after applying efficiencies. In the $0.5$--$1.5~\mathrm{keV}$ $E_{\mathrm{true}}$ slice, the simulated mean $(\mathrm{S1}c,\mathrm{S2}c)$ changes from approximately $(0.61,197)~\mathrm{phd}$ before efficiencies to $(3.59,334)~\mathrm{phd}$ after efficiencies, with a total efficiency of $0.9\%$. In the $1.5$--$2.5~\mathrm{keV}$ $E_{\mathrm{true}}$ slice, the corresponding means are $(1.28,430)~\mathrm{phd}$ and $(3.85,481)~\mathrm{phd}$, with an efficiency of $7.9\%$.
The three-fold S1 coincidence requirement preferentially retains upward S1 fluctuations, bringing the mean S1$c$ values after applying efficiencies closer together despite the different underlying responses. This explains why the lowest-energy contours cluster at similar S1$c$. These quantities are averages over finite-width $E_{\mathrm{true}}$ slices; applying efficiencies also changes the true-energy distribution within each interval.

For the $E_{\mathrm{true}}$ slices centered at $10$, $40$, and $74~\mathrm{keV}$, the total efficiencies exceed $90\%$, and the mean signals before and after applying efficiencies are much closer. Both S1$c$ and S2$c$ increase with recoil energy, and the contours follow the diagonal trend of the NR band. The UMS component also dominates the simulated high-energy-band population with $\mathrm{S2}c>9000~\mathrm{phd}$.

The overlap of the contours also affects their interpretation. The nominal enclosed probability of a contour refers to the estimated density of SS events after applying efficiencies in its defining $E_{\mathrm{true}}$ slice. The same region of $(\mathrm{S1}c,\mathrm{S2}c)$ space can also contain SS events from other true recoil energies, as well as UMS and accidental events. For example, the $1\sigma$ contour defined by the $0.5$--$1.5~\mathrm{keV}$ $E_{\mathrm{true}}$ slice contains approximately $2.8$ expected SS events from that defining interval, but approximately $19$ expected SS events when all true recoil energies are included. Adding the accidental components and UMS gives approximately $24$ expected model events, compared with $24$ DD data events. Thus, comparison with the observed count requires the total model within the contour; the defining $E_{\mathrm{true}}$ slice supplies only part of that expectation. 

The strong overlap of the low-energy $E_{\mathrm{true}}$ populations limits event-by-event recoil-energy discrimination near threshold. The low efficiencies further reduce the number of events available to constrain the yields. Statistical sensitivity to the yields is assessed separately by the yield-cutoff study in Fig.~\ref{fig:energy_sens}, which gives sensitivity thresholds of $2.2~\mathrm{keV}$ for the S1$c$ projection and $1.9~\mathrm{keV}$ for the S2$c$ projection. The gray efficiency curve in that figure includes both SS and UMS events, whereas Table~\ref{tab:dd_contour_counts} reports SS-only efficiencies. A small population of events survives the efficiencies below these sensitivity thresholds, where the constraints on the yields remain limited.

\begin{table*}[tbp]
\caption{Simulated single-scatter response and event counts for the $E_{\mathrm{true}}$ slices defining the contours in Fig.~\ref{fig:s1s2_contours}. For each $E_{\mathrm{true}}$ slice, the columns labeled ``Before efficiencies'' and ``After efficiencies'' give the mean S1$c$ and S2$c$ and the expected counts $N_{\rm before}$ and $N_{\rm after}$, respectively. These quantities refer to the full $E_{\mathrm{true}}$ slice, before restricting to the contour. The before-efficiency means are obtained from \textsc{nest} for the same parent SS sample, with the coincidence and S2 event thresholds disabled and zero-signal events retained. Detector response remains included. The after-efficiency means and counts include the final cut-acceptance weights. The total SS efficiency is $\epsilon=N_{\rm after}/N_{\rm before}$; it includes the three-fold coincidence requirement, reconstruction efficiencies, analysis cuts, and cut acceptances, and is conditional on the geometric requirements used to construct the simulated DD sample. The remaining columns give counts inside the corresponding $1\sigma$ contour, defined to enclose a nominal probability of $p=0.393$ in the estimated SS density. ``SS sim. ($E_{\mathrm{true}}$ slice)'' is the SS count from the defining $E_{\mathrm{true}}$ slice within that contour. The three ``Model components'' columns give the SS count from all true recoil energies, the UMS count from all true recoil energies, and the accidental-event count within the same contour. ``Total model'' is their sum and is compared with ``DD data''. Within the same contour, the SS events from the defining $E_{\mathrm{true}}$ slice form a subset of the SS events at all true recoil energies and are therefore already included in ``Total model''. All simulated counts are expected counts normalized to the DD data, with the same SS normalization before and after applying efficiencies. Component and total counts are rounded independently.}
\label{tab:dd_contour_counts}
\centering
\scriptsize
\setlength{\tabcolsep}{2.5pt}
\renewcommand{\arraystretch}{1.18}
\begin{ruledtabular}
\begin{tabular}{l|rrr|rrrr|r|rrr|rr}
& \multicolumn{7}{c|}{SS sim. ($E_{\mathrm{true}}$ slice)} & \multicolumn{6}{c}{$1\sigma$ contour ($p=0.393$)} \\
\cline{2-8}\cline{9-14}
& \multicolumn{3}{c|}{Before efficiencies} & \multicolumn{4}{c|}{After efficiencies} & \multicolumn{1}{c|}{Defining slice} & \multicolumn{3}{c|}{Model components} & \multicolumn{2}{c}{Model and data} \\
\cline{2-4}\cline{5-8}\cline{9-9}\cline{10-12}\cline{13-14}
\shortstack{Contour-defining\\$E_{\mathrm{true}}$ slice\\{[keV]}} &
\multicolumn{1}{c}{\shortstack{Mean\\S1$c$\\{[phd]}}} &
\multicolumn{1}{c}{\shortstack{Mean\\S2$c$\\{[phd]}}} &
\multicolumn{1}{c|}{\shortstack{Expected\\count\\$N_{\rm before}$}} &
\multicolumn{1}{c}{\shortstack{Mean\\S1$c$\\{[phd]}}} &
\multicolumn{1}{c}{\shortstack{Mean\\S2$c$\\{[phd]}}} &
\multicolumn{1}{c}{\shortstack{Expected\\count\\$N_{\rm after}$}} &
\multicolumn{1}{c|}{\shortstack{Total\\efficiency\\$\epsilon$ [\%]}} &
\multicolumn{1}{c|}{\shortstack{SS sim.\\($E_{\mathrm{true}}$ slice)}} &
\multicolumn{1}{c}{\shortstack{SS sim.\\(all $E_{\mathrm{true}}$)}} &
\multicolumn{1}{c}{\shortstack{UMS\\sim.}} &
\multicolumn{1}{c|}{\shortstack{Acc.}} &
\multicolumn{1}{c}{\shortstack{Total\\model}} &
\multicolumn{1}{c}{\shortstack{DD\\data}} \\
\hline
73.5--74.5 & 116.40 & 6705 & 235 & 114.99 & 6711 & 220 & 93.82 & 89 & 1059 & 32 & 0.0 & 1091 & 1019 \\
39.5--40.5 & 55.28 & 4717 & 89 & 54.17 & 4718 & 86 & 96.13 & 35 & 485 & 22 & 0.0 & 507 & 484 \\
9.5--10.5 & 9.47 & 1866 & 827 & 9.44 & 1867 & 757 & 91.56 & 302 & 1458 & 39 & 0.0 & 1497 & 1473 \\
2.5--3.5 & 2.03 & 653 & 993 & 4.19 & 675 & 207 & 20.83 & 83 & 288 & 4.1 & 2.1 & 294 & 281 \\
1.5--2.5 & 1.28 & 430 & 885 & 3.85 & 481 & 70 & 7.89 & 28 & 106 & 1.3 & 3.6 & 111 & 116 \\
0.5--1.5 & 0.61 & 197 & 730 & 3.59 & 334 & 6.7 & 0.92 & 2.8 & 19 & 0.2 & 4.7 & 24 & 24 \\
\end{tabular}
\end{ruledtabular}
\end{table*}

\FloatBarrier